\documentclass[%
footinbib,
superscriptaddress,
twocolumn,
amsmath,amssymb,
aps,
prl
]{revtex4-2}

\usepackage{ bbold }
\usepackage{graphicx}
\usepackage{dcolumn}
\usepackage{bm}
\usepackage[justification=RaggedRight]{caption}
\usepackage{subcaption}
\usepackage[utf8]{inputenc}
\usepackage[T1]{fontenc}
\usepackage{braket}
\usepackage{soul} 

\usepackage{dsfont}
\usepackage{xcolor}
\usepackage{tikz}
\usetikzlibrary{positioning}

\usepackage{xcolor}		
\usepackage[english]{babel} 

\newbox{\bigpicturebox}

\usepackage{csquotes}
\usepackage{hyperref}
\usepackage{bookmark}

\hypersetup{%
  breaklinks=true,
  colorlinks=true,
  linkcolor=black,
  filecolor=black,
  urlcolor=black,
  citecolor=black,
}

\begin{document}

\title{The role of nestedness and saturating feedback in bipartite ecological systems}

\author{Nirbhay Patil}
\affiliation{Laboratoire Matière et Systèmes Complexes (MSC), Université Paris Cité $\&$ CNRS (UMR 7057), 75013 Paris, France}

\author{Ada Altieri}
\affiliation{Laboratoire Matière et Systèmes Complexes (MSC), Université Paris Cité $\&$ CNRS (UMR 7057), 75013 Paris, France}

\begin{abstract}
Large ecosystems balance competition and cooperation, yet standard generalized Lotka--Volterra models make mutualism destabilizing by amplifying disorder and driving unbounded growth. We show that Monod-like saturation resolves this paradox: dynamical mean-field theory and random-matrix analysis reveal a broader stable phase and enhanced survival. Network architecture provides a second control mechanism, but nestedness offers no intrinsic stability advantage. Instead, it is a byproduct of degree distributions with high connectivity necessary for stability.
\end{abstract}

\maketitle

Since the seminal works of Lotka and Volterra~\cite{lotka1920,volterra1927},
Lotka--Volterra dynamics has become a paradigmatic framework for describing
interacting ecological communities. A central question, dating back to
May's complexity--stability argument~\cite{may1972,may1976, allesina2012stability, hatton2024diversity}, is how the
collective behavior of a community changes with the number of species and
the heterogeneity of their interactions. Random-matrix theory and Dynamical Mean-Field Theory have subsequently provided a detailed
characterization of large random ecosystems and of the transitions
between qualitatively distinct dynamical regimes~\cite{Bunin2017,galla2018dynamically,Roy2020,Altieri2020dynamical}. In particular, Generalized Lotka--Volterra models with random interactions display three robust phases: a single-equilibrium phase, where all initial conditions converge to the same fixed point; a multiple-attractor phase where the number of fixed points turns out to be exponential in the system size \cite{Altieri2021, ros2023generalized}; and an unbounded growth phase, where the dynamics cease to admit a finite stable stationary state \cite{Bunin2017,Altieri2020dynamical}. 
Beyond their theoretical relevance, multistability and chaotic dynamics may have important ecological consequences, enabling communities with the same interaction structure to display pronounced history dependence and to transition among alternative states, as observed in recent experiments on microbial communities \cite{gore2025transition, abreu2020, aguade2024taxonomy}.

Their emergence is controlled by the statistics of the interaction matrix. Increasing interaction heterogeneity destabilizes the unique equilibrium, eventually leading to multistability or unbounded growth. Likewise, a negative mean interaction -- corresponding in our convention to an excess of mutualistic over competitive effects -- promotes runaway dynamics. This may appear counterintuitive, since mutualism is generally associated with ecological benefit and functional robustness. In microbial communities, for example, cross-feeding can support metabolic diversity, while host-microbe interactions may enhance colonization resistance, immune regulation, and nutrient processing \cite{chacon2021ecology, caballero2023}. 

This apparent contradiction points to a fundamental limitation of the
standard linear framework. In real ecosystems, the benefit received from
mutualistic partners cannot increase indefinitely. Metabolic pathways
saturate, resources become limiting, and the benefit supplied by one
species may plateau once another ecological or physiological constraint
becomes limiting. This is particularly natural in microbial communities,
where effective interactions are often mediated by finite supplies of
metabolites, nutrients, or host-controlled resources \cite{tu2019}. A realistic theory
must therefore account not only for the sign and heterogeneity of the
interactions, but also for the saturation of positive feedback. Motivated
by this limitation, several extensions of Lotka--Volterra dynamics have
introduced nonlinear functional responses, ecological structure, spatial
dependence, and environmental fluctuations~\cite{baron2020dispersal,garcia2024interactions, altieri2022effects}.

A second central issue concerns the architecture of the interaction network \cite{montoya2006ecological,pascual2005ecological}. Mutualistic communities, particularly plant--pollinator systems, often display nontrivial structural organization, commonly characterized in terms of nestedness and modularity \cite{bascompte2003nested,bastolla2009architecture,olesen2007modularity,fortuna2010nestedness, pichon2024}. In a nested bipartite network, the interaction partners of specialists tend to form subsets of those of increasingly generalist species. Nestedness, therefore, encodes correlations between species degrees and interaction patterns that are absent from unstructured random-network models. Determining how such structural correlations affect collective dynamics remains a central problem in ecological and network theory \cite{alves2019}.

\begin{figure*}[t]
    \centering\hspace{-5mm}
    \includegraphics[height=3.6cm]{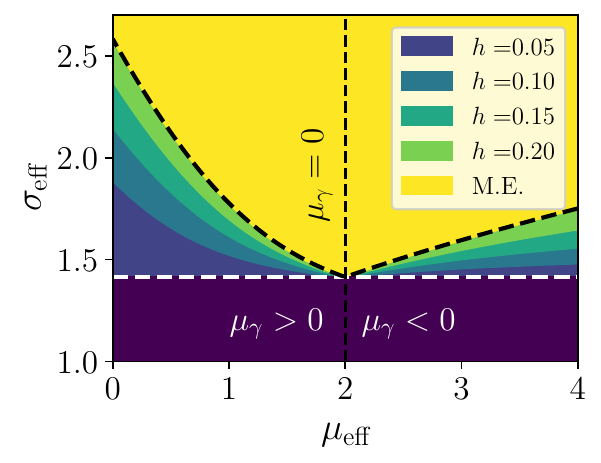}
    \includegraphics[height=3.7cm]{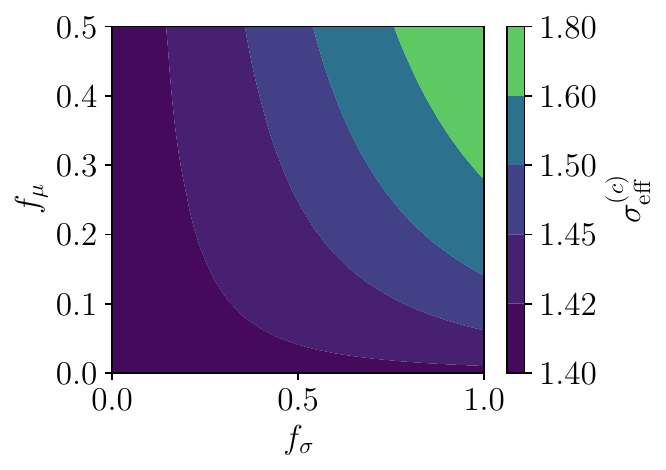}
    \includegraphics[height=4cm]{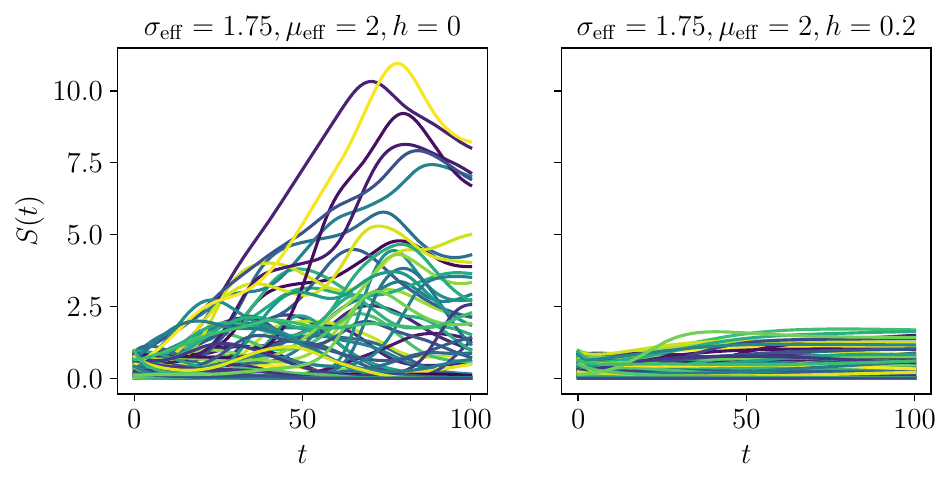}
    \caption{The phase boundary between the Unique Fixed Point and Multiple Attractor phases moves upwards as we increase $h$. Left plot: the axes represent the effective mean and variances $\mu_{\rm eff}=\mu_\beta-\mu_\gamma$ and $\sigma_{\rm eff}^2=\sigma_\beta^2+\sigma_\gamma^2$. We see that if there is no saturation (represented by the dashed white line at $h=0$), the transition occurs at $\sigma_{\rm eff}^{(c)}=\sqrt2$ as expected from a fully connected graph with no correlation in the disorder. Keeping $\mu_\beta=2$, the phase transition to the multiple-attractor regime requires higher values of the disorder variance, regardless of the sign of the added non-linear bipartite network. Each region shows the value of $h$ at which it exhibits a single fixed point, dark purple being stable for all values of $h$ while yellow beyond the dashed black boundary showcases multiple attractors for all values of $h\leq0.2$. The second plot shows the exact position of the critical line $\sigma_{\rm eff}^{(c)}$ as a function of  $f_\mu=\mu_\gamma/\mu_{\rm eff}$ and $f_\sigma=\sigma_\gamma/\sigma_{\rm eff}$. In the right panels, the time series highlight the markedly different phases for different values of $h$, at fixed $\mu_\beta=4$, $\mu_\gamma=2$, $\sigma_\beta=0.5$, $\sigma_\gamma=1.68$.}
    \label{fig:singleeq_bound}
\end{figure*}


The literature on plant--pollinator systems \cite{bastolla2009architecture,saavedra2011strong} commonly introduces saturating mutualistic responses to prevent unbounded growth \cite{wright1989simple}. Here, we study a generalized Lotka--Volterra (GLV) model with random Gaussian interactions, in which mutualistic effects are mediated by a structured bipartite network and saturate at large total input. In contrast with recent nonlinear extensions of fully connected GLV systems~\cite{sidhom2020ecological, zenari2025generalized}, our framework allows us to disentangle nonlinear feedback from degree heterogeneity, degree correlations, and nestedness. We show that the stabilizing effect of nonlinear mutualism depends not only on the overall interaction strength, but also on how disorder is partitioned between the competitive and mutualistic sectors and on the topology of the bipartite network.


By employing a Dynamical Mean-Field Theory (DMFT) formalism, we identify a stabilization mechanism
that is absent from unstructured linear random-interaction ensembles.
The phase behavior depends not only on the total interaction
heterogeneity, but also on the fraction of disorder carried by the
saturating mutualistic sector and on the degree structure of the
bipartite graph. Saturation suppresses positive feedback, whereas
degree heterogeneity and disassortativity can counteract this benefit.
Consequently, nestedness does not constitute an intrinsically stabilizing
property: its dynamical effect must be assessed relative to the degree
sequence from which it emerges.

Notably, we consider two interacting sectors, with predominantly competitive interactions within each and mutualistic interactions across the two groups. Plant--pollinator communities provide the canonical example, although the same structure may also describe interacting microbial guilds or effective couplings between microorganisms and exchanged metabolites, nutrients, or host-derived factors.  

We denote by
\(P_i\) the abundances of species in the first group --- with $i=1,\ldots,N_P$ ---
and by \(A_i\) those in the second group --- where $i=1,\ldots,N_A$. Their
dynamics are eventually captured by:
\begin{align}
    \begin{aligned}
        \dot P_i&=P_i\left(1-P_i-\sum_j\beta_{ij}^PP_j+\frac{\sum_jC_{ji}\gamma^P_{ij}A_j}{1+|h^P\sum_kC_{ki}\gamma^P_{ik}A_k|}\right) \ , \\
          \dot A_i&=A_i\left(1-A_i-\sum_j\beta_{ij}^AA_j+\frac{\sum_jC_{ij}\gamma^A_{ij}P_j}{1+|h^A\sum_kC_{ik}\gamma^A_{ik}P_k|}\right) \ .
    \end{aligned}
\end{align}
Here the coefficients $\beta_{ij}^{A(/P)}$ represent the competitive interactions within animals (or plants), while the coefficients $\gamma_{ij}^{A(/P)}$ reflect the mutualistic effect that animals receive via plants (or vice versa). The elements of the adjacency matrix $C_{ij}$ take on value $1$ if plant $j$ is connected to animal $i$ and $0$ otherwise. 
We then consider a nonlinear mutualistic term of Monod-like form, where the total positive input received by a species (or a group) is bounded by a saturation parameter, $h$ controlling the strength of such a  saturation. 


In the large-dimensional limit, where the numbers of degrees of freedom in the two interacting groups (both denoted by $N$) tend to infinity, DMFT provides a powerful tool to analyze the effective dynamics of a representative element of the system \cite{Altieri2020dynamical, Roy2020, galla2024generating, cugliandolo2023recent}. As commonly assumed in disordered systems, we take these coefficients to be random variables whose means and variances scale as $1/N$. More specifically, the competitive coefficients $\beta_{ij}$ are assumed to have mean $\mu_\beta/N$ and variance $\sigma_\beta^2/N$, and similarly for the mutualistic coefficients ($\gamma$), with mean $\mu_\gamma/C$ and variance $\sigma_\gamma^2/C$, where $C$ is the average number of plant-animal connections. For simplicity, we assume that the two groups are characterized by the same statistics.
The competitive interaction coefficients are assumed to have a covariance $\langle\beta_{ij}\beta_{ji}\rangle_c=c_\beta\sigma^2_\beta/N$. On the other hand, the covariance of the mutualistic terms is specifically between the coefficients determining the effect of plant $i$ on animal $j$ ($\gamma_{ji}^A$), and animal $j$ on plant $i$ ($\gamma_{ij}^P$), giving $\langle \gamma_{ij}^A\gamma_{ji}^P\rangle_c=c_\gamma\sigma^2_\gamma/C  $. 

Furthermore, to be as general as possible and model more realistic scenarios, the adjacency matrix $C_{ij}$ is assumed to have a degree distribution $\rho(v_A)$ for animals and $\rho(v_P)$ for plants, giving the probability of each node to be connected to $vN$ other nodes. Under these assumptions, applying DMFT analysis to the system of equations reduces them to effective equations for a representative plant (or animal) species of connectivity $v$, giving

\begin{align}
    \begin{aligned}
        \dot P_v(t)&=P_v(t)\left(1-P_v(t)-\beta_{{\rm eff}}^{(P)}(t)+\frac{\gamma_{{\rm eff} \;(v)}^{(P)}(t)}{1+|h\gamma_{{\rm eff} \;(v)}^{(P)}(t)| }\right),\\
        \beta_{{\rm eff}}^{(P)}&=\mu_\beta m_P(t)+\eta_P(t)+c_\beta\sigma_\beta^2\int dt'K_P(t',t)P_v(t'),\\
        \gamma_{{\rm eff} \;(v)}^{(P)}&=\mu_\gamma B_v(t)+\xi_v(t)+c_\gamma\sigma_\gamma\int dt'y_v(t',t)A_v(t'),
    \end{aligned}
\end{align}
and similarly for a representative animal species. The fields \(\eta(t)\) and \(\xi_{v}(t)\) are Gaussian effective noises generated by the
quenched interaction disorder, whereas \(K\) and \(y_{v}\) are
retarded response kernels encoding the feedback of a perturbation at an
earlier time on the subsequent dynamics. These quantities are determined
self-consistently by averaging over the effective single-species process
and over the corresponding degree distributions. Their explicit
expressions, obtained by extending the formalism of Ref.~\cite{patil2026dynamical} to
directed bipartite networks, are reported in the Appendix. We also show that degree correlations in the form of nestedness can be added to this formalism as well, the conclusions of which are presented in the End Matter.

\begin{figure}
    \centering\hspace{-5mm}
    \includegraphics[width=\linewidth]{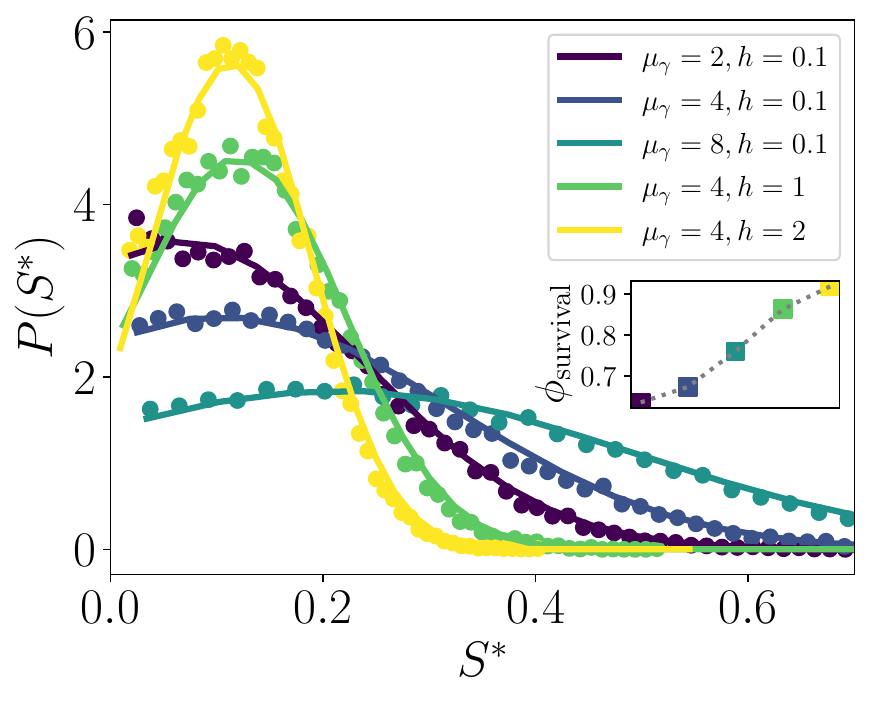}
    \caption{Theoretical predictions of the stationary species abundance distributions overlaid on simulation results represented by points, with $\mu_\beta=10$, $\sigma_\beta=0.5$, and $\sigma_\gamma=1$. Notice that both parameters have a beneficial effect on the survival probability, as shown in the inset plot. Results for fully connected networks.}
    \label{fig:statdistcomparison}
\end{figure}
We show that saturating mutualism qualitatively reshapes the phase diagram: it suppresses runaway growth, stabilizes finite-abundance stationary states, and modifies the onset of multiple-attractor behavior. Our results provide a statistical-physics framework for understanding how nonlinear mutualistic feedbacks can stabilize complex ecosystems, with direct relevance to microbial communities, resource-mediated interactions, and more general bipartite ecological networks.

Furthermore, treating the degree distribution $\rho(v)$ as an independent functional parameter, rather than restricting the analysis to fully connected networks, allows us to incorporate structural heterogeneity, degree correlations, and nestedness within the same framework. As evidenced in a number of empirical studies \cite{montoya2002small,jordano2003invariant,mossa2002truncation}, ecological networks --- with a particular focus on plant--pollinator systems --- are highly heterogeneous, approaching scale-free behaviour with fat-tailed degree distributions. Accounting for such heterogeneity effects is an essential step toward connecting theory with empirical network structure, a feature that is largely absent from existing studies for reasons of analytical tractability.

First, we can notice that in the linear fully connected limit, i.e. for $C_{ij}=1$ and $h=0$, the competitive and mutualistic contributions combine into an effective random interaction matrix, the stability of which can be calculated in several ways \cite{Sompolinsky1982,poley2024eigenvalue,patil2024spectral}. The resulting GLV-like dynamic is therefore captured by an effective mean and an effective variance of the disorder with $\mu_{\rm eff}=\mu_\beta-\mu_\gamma$ and $\sigma_{\rm eff}^2=\sigma_\beta^2+\sigma_\gamma^2$. 
As shown in Ref.~\cite{Bunin2017}, in the absence of correlations, the single-equilibrium solution becomes unstable above the critical value $\sigma_{\mathrm{eff}}^2=2$, leading to a multiple-attractor phase, as expected for the standard GLV model. As the effective mean interaction becomes more mutualistic (i.e. negative) and exhibits larger variance, the system eventually enters an unbounded-growth phase. Thus, in the purely linear limit, mutualistic interactions have a twofold destabilizing effect: they shift the effective mean toward positive feedback and increase the total strength of the disorder. As a consequence, they reduce the stability of the ecosystem and lower survival, contrary to the naive expectation that mutualistic interactions should generically enhance persistence.

Figure~\ref{fig:singleeq_bound} shows that this conclusion is qualitatively overturned once nonlinear saturation becomes relevant. In the linear regime, a stronger mutualistic sector increases the effective disorder and destabilizes the ecosystem. When positive feedback saturates, however, the same interactions suppress runaway growth and damp dynamical fluctuations, thereby shifting the instability threshold to larger disorder strengths. Stability is therefore controlled not by the total mean and variance alone, but by the fraction of each carried by the nonlinear mutualistic sector. To first order in the saturation parameter $h$, the critical condition for the instability of the single-equilibrium phase becomes:
\begin{align}
    \sigma_{\rm eff}^2\left[1-4h\left(\frac{\sigma_\gamma}{\sigma_{\rm eff}}\right)^2\frac{\mu_\gamma}{\mu_{\rm eff}}\right]=2.
\end{align}

This outcome highlights the mixed nature of cooperative interactions in the model. On the one hand, mutualism increases the effective disorder and shifts the effective mean toward positive feedback, both of which destabilize the ecosystem in the corresponding linear model. On the other hand, the saturation of the interspecific effects regularizes this positive feedback and prevents the divergence of the solutions. 

Such a mechanism is visible in Fig.~\ref{fig:statdistcomparison}, where we compare the stationary abundance distributions predicted by DMFT with direct numerical simulations.
Empirical microbial communities display robust non-Gaussian abundance statistics: species-level abundance fluctuations are Gamma distributed, while mean abundances across species are approximately Lognormal~\cite{grilli2020macroecological}. Our results suggest that nonlinear mutualistic feedbacks provide a natural dynamical mechanism for such skewed abundance patterns. Remarkably, unlike standard random GLV models, which typically yield truncated Gaussian stationary abundance distributions in the single-equilibrium phase, saturating mutualism reshapes the abundance distribution and simultaneously increases survival, thereby linking empirically relevant SAD shapes to ecological persistence.
Increasing either the mean strength of the mutualistic interactions or the saturation parameter $h$ leads to higher survival probabilities. Nonlinear saturation also allows the system to sustain much larger effective variances without entering the unbounded-growth phase. Overall, our findings reveal that under the assumption that mutualistic effects saturate, ecosystems are stable for a wider range of variances of the interaction coefficients. In addition, due to the saturation term, no amount of disorder in the ``plant to animal'' interactions can cause the system to reach a state of unbounded growth.

\begin{figure}
    \centering
    \includegraphics[scale=0.5]{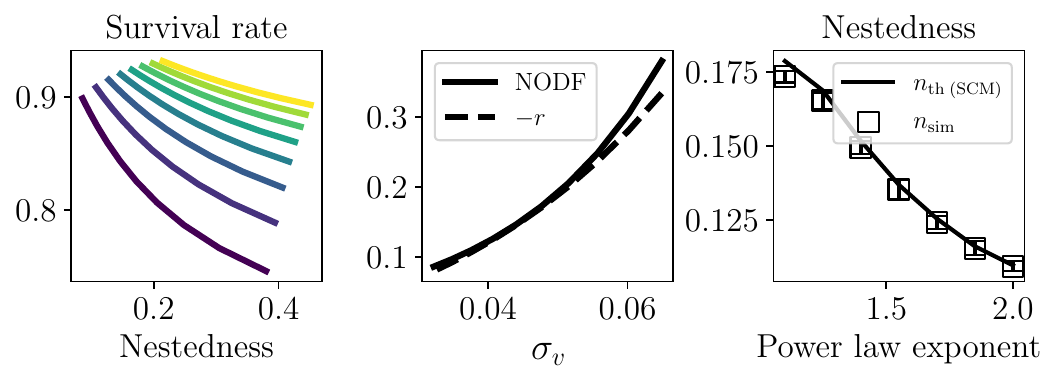}
    \caption{The survival rate of species plotted alongside the nestedness for different values of the average connectance varying from 0.01 (deep purple) to 0.1 (yellow), taking $\mu_\beta=0.1,\mu_\gamma=0.15,\sigma_\beta=0.1,\sigma_\gamma=0.5$ and $h=0.01$. For each value of the average connectivity chosen, we take a power-law distributed graph, and calculate its nestedness, disassortativity (negative of the assortativity $r$), and degree heterogeneity. We are unable to observe any benefits provided by the nestedness of a graph, as it isn't possible to isolate this feature from the degree heterogeneity which makes the system less stable. The right most plot shows how a more fat-tailed power law (i.e. a lower power law exponent) automatically has higher nestedness in simulated graphs, which matches very well the predicted theoretical nestedness of these graphs from our theory using the soft configurational model.}
    \label{fig:degreedistplot}
\end{figure}

We also investigated whether network architecture can further modulate the stabilizing effect of saturating mutualism. More precisely, Fig.~\ref{fig:degreedistplot} shows the survival probability as a function of the nestedness overlap with decreasing fill (NODF) \cite{almeida2008consistent,payrato2020measuring} for bipartite networks with heterogeneous degree distributions. Although nestedness is often expected to enhance stability in mutualistic systems \cite{bastolla2009architecture}, our results show that its effect cannot be isolated straightforwardly in random heterogeneous networks. The first plot in Fig. \ref{fig:degreedistplot} shows this ambivalence of the system to the NODF -- either increasing the average degree $C$ or reducing the power law exponent $\alpha$ both lead to an increase in nestedness and degree heterogeneity, while the former leads to higher survival and the latter lower. This could possibly be explained by the fact that the stability of the system drops heavily for lower average degrees \cite{patil2026dynamical} -- thus forcing any stable system to have a high connectivity, and consequently be nested due to structural constraints.

\begin{figure}
    \centering
    \includegraphics[scale=0.5]{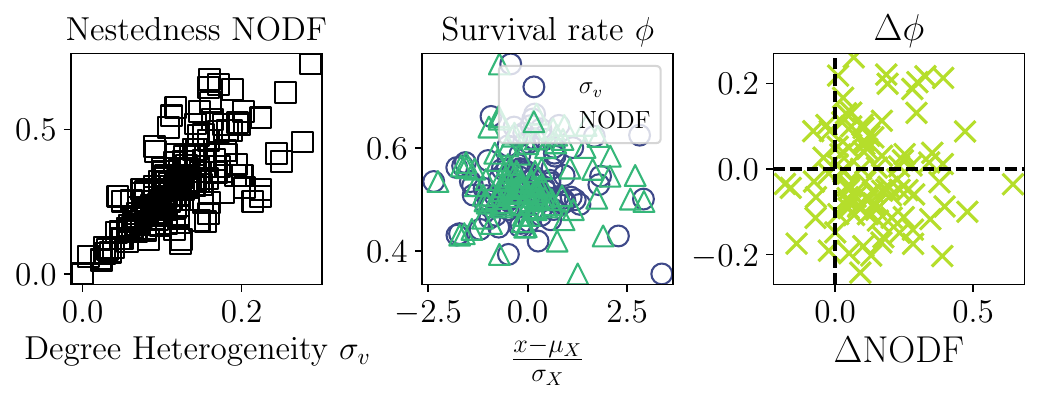}
    \caption{For real networks from the Web of Life database, we show that nestedness is strongly correlated with degree heterogeneity. Consistent with our theoretical predictions, species survival shows no systematic dependence on either degree heterogeneity or rescaled, demeaned nestedness. Moreover, degree-preserving randomized networks can differ substantially in nestedness without exhibiting corresponding changes in survival. Among networks that are more nested than expected from their degree sequence, 56$\%$ actually display lower survival than their randomized counterparts.
    }
    \label{fig:dataplot}
\end{figure}

Building on the observation that nestedness can be a statistical by-product of the degree sequence~\cite{payrato2019,jonhson2013factors} due to structural cut-offs \cite{boguna2002epidemic,catanzaro2005generation,zhou2007structural,patil2026dynamical}, we show that this structural degeneracy has dynamical consequences: degree-induced nestedness is not an independent source of stability, but is entangled with heterogeneity-driven fluctuations that can destabilize nonlinear mutualistic ecosystems.
In the ensembles considered here, increasing nestedness is strongly correlated with increasing degree heterogeneity and disassortativity, both of which tend to destabilize the dynamics. Consequently, we do not observe a direct stabilizing effect of nestedness alone. In fact, if one assumes that either of the two types of species dynamics are much faster than the other -- leading them to be always in equilibrium, we find that the effective competitive interactions (Eq. \ref{eq:separationscaleseffectivestats}) within the slow species are modified by terms related to the nestedness in a way that also reduce the mean while increasing the variance, unequivocally leading to higher instability and lower species biodiversity. The same pattern repeats when we introduce nestedness to the HDMFT framework, finding that it results in additional noise and response terms that depend on the difference between the actual nestedness of the network and the expected nestedness of a network with a given degree distribution (eq. \ref{eq:DFMT_NODF}).

Rather, the dominant structural feature appears to be the organization of the degree sequence itself. To disentangle these effects, we compare the measured nestedness with the prediction of the soft configuration model, a random bipartite graph ensemble in which the degrees are fixed on average, while being as random as possible, thus ensuring minimum correlations. The agreement between simulations and this null-model prediction shows that the observed nestedness is largely induced by the fat-tailed degree distribution, rather than by any architectural principle. In Fig. \ref{fig:dataplot}, we calculate the properties of real networks taken from the Web of Life dataset. We find a strong correlation between nestedness and degree heterogeneity as expected, but neither property seems to be correlated to the survival rates of Lotka-Volterra systems simulated on top of these networks. In the rightmost plot, we show the difference in survival rate between a simulation on the exact network and a generic network with the same degree distribution, as a function of the difference between the nestedness of the two. We find that networks that are more nested than expected do not have higher survival rates, with more than half the cases studied having reduced biodiversity in these cases. 


In conclusion, our results point out that the stability of mutualistic ecosystems cannot be inferred from interaction signs alone. Nonlinear saturation removes the unphysical runaway growth of linear GLV dynamics and enlarges the stable phase, but degree heterogeneity can counteract this benefit by amplifying fluctuations. Nestedness, in turn, is largely constrained by the degree sequence and does not constitute an independent source of stability. These findings shift the stability problem from the mere presence of cooperation to the way cooperative feedback is distributed across the network, challenging the common view that nestedness is intrinsically beneficial and identifying the joint statistics of saturation, disorder, and degree heterogeneity as the relevant control parameters.

\subsection*{Acknowledgements}
AA and NP acknowledge the support of ANR JCJC ‘SIDECAR’ ANR-23-CE30-0012-01.
 This research was supported in part by grant NSF PHY-2309135 to the Kavli Institute for Theoretical Physics (KITP). We would also like to thank valuable feedback from Guim Aguade, Fabian Aguirre-L\'opez,  Thibaut Arnoux de Pierry, and Samir Suweis. 

\bibliography{Bibliog}

\appendix

 \section{End Matter}

 \subsection{A. Bipartite self-consistency equations}
 The effective dynamics of a representative plant/animals species with connectivity $v$ can be obtained by performing the heterogeneous DMFT, i.e. a refined formalism that allows us to extend standard DMFT to systems in which agents or species are not statistically equivalent by basically labeling the different effective processes through degree, spatial position, interaction class, etc.\cite{dorogovtsev2008critical,pastor2001epidemic,park2024incorporating,aguirre2024heterogeneous,poley2025interaction}. We extend the generic results presented in \cite{patil2026dynamical} for undirected networks with arbitrary degree distributions to include bipartite or directed networks, which give us expressions for the population dynamics of a representative plant (or animal) species of fixed degree $P_v$ ($A_v$):
\begin{align}
    \begin{aligned}
        \dot P_v(t)&=P_v(t)\left(1-P_v(t)-\beta_{{\rm eff}}^{(P)}(t)+\frac{\gamma_{{\rm eff} \;(v)}^{(P)}(t)}{1+|h\gamma_{{\rm eff} \;(v)}^{(P)}(t)| }\right),\\
         \dot A_v(t)&=A_v(t)\left(1-A_v(t)-\beta_{{\rm eff}}^{(A)}(t)+\frac{\gamma_{{\rm eff} \;(v)}^{(A)}(t)}{1+|h\gamma_{{\rm eff} \;(v)}^{(A)}(t)| }\right).
         \end{aligned}\label{eq:fulleffequations}
         \end{align}
         
        In this representation, the mutualistic and competitive interactions get condensed into the effective terms $\beta_{\rm eff}$ and $\gamma_{\rm eff}$, given by
        
        \begin{align}\begin{aligned}
        \beta_{{\rm eff}}^{(P)}&=\mu_\beta m_P(t)+\eta_P(t)+c_\beta\sigma_\beta^2\int dt'K_P(t',t)P_v(t'),\\
        \gamma_{{\rm eff} \;(v)}^{(P)}&=\mu_\gamma B_v(t)+\xi_v(t)+c_\gamma\sigma_\gamma^2\int dt'y_v(t',t)A_v(t'),\\
        \beta_{{\rm eff}}^{(A)}&=\mu_\beta m_A(t)+\eta_A(t)+c_\beta\sigma_\beta^2\int dt'K_A(t',t)A_v(t'),\\
        \gamma_{{\rm eff} \;(v)}^{(A)}&=\mu_\gamma D_v(t)+\chi_v(t)+c_\gamma\sigma_\gamma^2\int dt'z_v(t',t)P_v(t').
    \end{aligned}
\end{align}
 The parameters above encode the information coming from the species populations and interaction coefficients, the three terms each in order being - \textit{(a)} the average interaction strength, \textit{(b)} a coloured noise encoding the disorder in the interactions with $ \langle \eta_P(t)\eta_P(t')\rangle=q_P(t,t')$ and $ \langle \eta_A(t)\eta_A(t')\rangle=q_A(t,t')$, and \textit{(c)} a feedback kernel arising from correlations in the interaction matrix. The various parts of the effective competitions are defined as follows,
\begin{align}
    \begin{aligned}
        m_P(t)&=\Big\langle P_v(t)\Big\rangle\qquad&  m_A(t)&=\Big\langle A_v(t)\Big\rangle\\
       q_A&=\Big\langle P_v(t)P_v(t')\Big\rangle\qquad&q_P&=\Big\langle A_v(t)A_v(t')\Big\rangle\\
        K_P(t,t')&=\Big\langle \frac{\delta P_v(t')}{\delta \eta_P(t)}\Big\rangle\qquad&
        K_A(t,t')&=\Big\langle \frac{\delta A_v(t')}{\delta \eta_A(t)}\Big\rangle,
    \end{aligned}
\end{align} where the averages are taken over degrees and different trajectories corresponding to all possible noises. The degree-dependent effective mutualism terms have slightly more complicated expressions for the order parameters, with
 
\begin{align}
    \begin{aligned}
        B_v(t)&=\frac1{\langle v_A\rangle}\int dv_A \rho(v_A)\Big\langle P(v_A,v_P)A_v(t)\Big\rangle,\\
        \langle \xi_v(t)\xi_v(t')\rangle&=\frac{\sigma_P^2}{\langle v_A\rangle}\int dv_A \rho(v_A)\Big\langle P(v_A,v_P)A_v(t)A_v(t')\Big\rangle,\\
        y_v(t,t')&=\frac1{\langle v_a\rangle}\int dv_A\rho(v_A)\Big\langle P(v_A,v_P)\frac{\delta A_v(t)}{\delta\xi_v(t')}\Big\rangle.\\
        D_v(t)&=\frac1{\langle v_P\rangle}\int dv_P \rho(v_P)\Big\langle P(v_A,v_P)P_v(t)\Big\rangle,\\
        \langle \chi_v(t)\chi_v(t')\rangle&=\frac{\sigma_A^2}{\langle v_P\rangle}\int dv_P \rho(v_P)\Big\langle P(v_A,v_P)P_v(t)P_v(t')\Big\rangle,\\
        z_v(t,t')&=\frac1{\langle v_P\rangle}\int dv_P\rho(v_P)\Big\langle P(v_A,v_P)\frac{\delta P_v(t)}{\delta\chi_v(t')}\Big\rangle.
    \end{aligned}\label{eq:effmutualisticequations}
\end{align}

Normally, the probability of connection of two nodes of degrees $v_A$ and $v_P$ is assumed to be given by the Chung-Lu model \cite{chung2002connected,poley2025interaction,park2024incorporating,aguirre2024heterogeneous} with $P(v_A,v_P)=v_Av_P/\sqrt{c_Ac_P}$, where $c$ is the average connectivity of animals or plants. However, as discussed in \cite{patil2026dynamical}, this misses some spurious correlations arising in the degrees due to structural constraints in scale-free networks. To account for such highly heterogeneous networks which are ubiquitous in ecological systems, we consider the soft-configuration model, which is the in essence the most random way to connect nodes in a graph to obtain a given degree distribution, or the maximum entropy ensemble. The probability of connection of nodes in this ensemble is
\begin{align}
    P(v_A,v_P)=\frac{\tilde v_A(v_A)\tilde v_P(v_P)}{\tilde v_A(v_A)\tilde v_P(v_P)+\sqrt{c_Ac_P}}.
\end{align}
Here the variables $\tilde v$ are ``hidden degrees'' assigned to each node \cite{park2004statistical,park2003origin}, that can be computed by realising that the actual degrees of the system are then given by
\begin{align}
    \begin{aligned}
v_A&=\int d\tilde v_P\tilde \rho(\tilde v_P)P(\tilde v_A\tilde v_P)    \\
v_P&=\int d\tilde v_A\tilde \rho(\tilde v_A)P(\tilde v_A\tilde v_P)  \ .
    \end{aligned}
\end{align}

The hidden degrees can then be calculated as in \cite{patil2026dynamical} by iterating over the expressions
\begin{align}
\begin{aligned}
\tilde v_A&=\frac{v_A}{\int d\tilde v_P\tilde \rho(\tilde v_P)\frac{\tilde v_P}{\tilde v_A\tilde v_P+\sqrt{c_Ac_P}}}\\
\tilde v_P&=\frac{v_P}{\int d\tilde v_A\tilde \rho(\tilde v_A)\frac{\tilde v_A}{\tilde v_A\tilde v_P+\sqrt{c_Ac_P}}}.
\end{aligned}
\end{align}
This lets us obtain the true probability of connection of nodes that includes the forced correlations, and can be inserted back into Eq. \ref{eq:effmutualisticequations} to obtain the effective interactions in heterogeneous networks. 
\begin{figure*}
    \centering
    \includegraphics[width=0.7\linewidth]{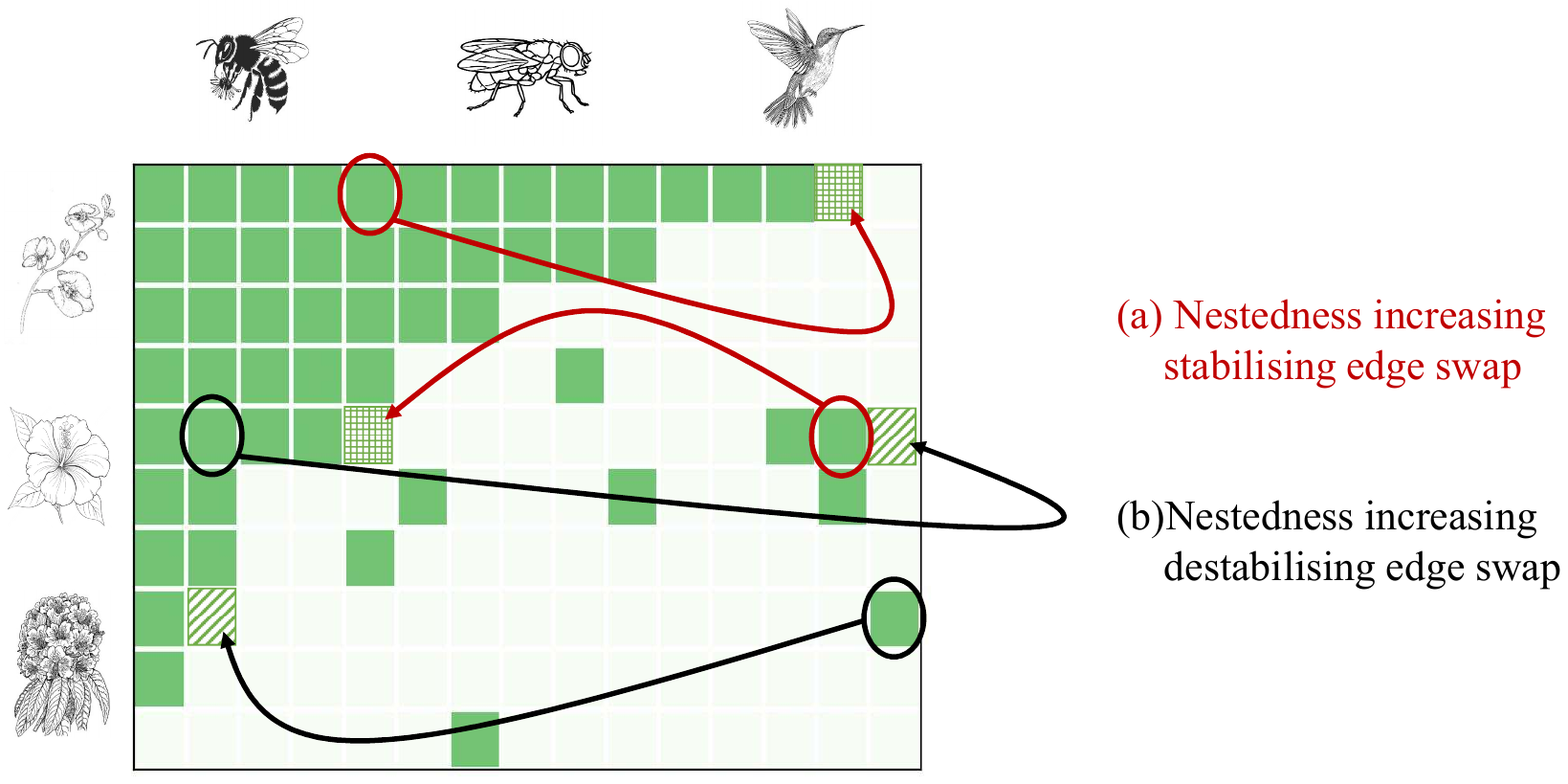}
    \caption{Here we construct an adjacency matrix for a bipartite network, which has a significant amount of nestedness. Trying out degree-preserving edge swaps, we find that only 12.3\% of them in this particular case increase the fractional nestedness defined in Eq. \ref{eq:NODFdef} while also reducing the overall overlap, thus stabilizing the system. On the other hand, 48\% of edge swaps lead to a higher NODF but also increase the total overlap, hence making the system less stable. We draw a schematic of two different swaps, one in red that increases the nestedness and makes the system more stable, and one in black that destabilizes the system. As the swaps are degree-preserving, the \textit{expected} nestedness of all of these networks is the same, and thus we are only changing the objects $f(ik,jk)$ that denote the excess overlaps for specific indices.}
    \label{fig:exmaplediagram}
\end{figure*}

\subsection{B. Reassessing the role of nestedness}

In the process leading up to the expression for the effective dynamics of a species of a certain degree given in Eq. \ref{eq:fulleffequations}, we arrive at a point where
one needs to calculate the average 
\begin{align}
    \begin{aligned}     \overline{\prod_{ij}\exp\Biggr\lbrace-iC_{ij}\left(\hat a_i\gamma^{(P)}_{ij}A_j+\hat p _j\gamma^{(A)}_{ji}P_i\right)\Bigg\rbrace}.
    \end{aligned}
\end{align} Now expanding this over all indices $i$ and $j$ would give many overlap terms for the adjacency matrix, of which the second-order overlap $C_{ki}C_{kj}$ is the object essential to having nestedness in networks. The measure NODF that we use is defined as
\begin{align}
    \begin{aligned}
        {\rm NODF}=\sum_{ij}\Theta(v_j-v_i)\frac{\sum_kC_{ki}C_{kj}}{v_i}.\label{eq:NODFdef}
    \end{aligned}
\end{align}
Assuming a separation of timescales in which animal species relax to equilibrium much faster than plants, the adiabatic approximation yields the following effective competitive interactions among the (slower) plant species:
\begin{align}
    \begin{aligned}
\beta'^{(P)}_{ij}&=\beta^{(P)}_{ij}-\gamma^{(P)}_{ij}C_{ji}-\sum_{k}\gamma^{(P)}_{ik}C_{ki}C_{kj}\gamma^{(A)}_{kj}\\&\qquad+\sum_l\gamma^{(P)}_{il}C_{li}\beta^{(A)}_{lj}+\sum_{kl}\gamma^{(P)}_{il}C_{li}\beta^{(A)}_{lk}C_{kj}\gamma^{(A)}_{kj} \ .
    \end{aligned}
\end{align}
Accordingly, we find that both node degrees and pairwise overlaps enter the statistics of the effective interaction matrix, leading to
\begin{align}
    \begin{aligned}
        \mu_{ij}'&=\mu_\beta+a(v_i,v_j)-cf_{ij}{\rm min}(v_i,v_j),\\
(\sigma_{ij}')^2&=\sigma_\beta^2+b(v_i,v_j)+df_{ij}{\rm min}(v_i,v_j),
    \end{aligned}\label{eq:separationscaleseffectivestats}
\end{align}
where $f_{ij}=\frac{\sum_kC_{ki}C_{kj}}{{\rm min}(v_i,v_j)}$ is the fractional overlap between partners of nodes $i$ and $j$. In particular, $f_{ij}=1$ when all partners of a less-connected node are also partners of the other one. The NODF as per its definition is then given by the sum $F=\sum_{ij}f_{ij}$. Because the entries of $f$ are constrained by the underlying bipartite network, increasing $F$ does not uniquely determine the sum
\begin{equation}
    X = \sum_{ij} f_{ij}\min(v_i,v_j) \ ,
\end{equation}
which controls the overall variance of the effective competition matrix. Although network rearrangements can, in principle, increase $F$ while decreasing $X$, a generic increase in nestedness need not do so. Therefore, higher NODF does not, by itself, imply greater stability or biodiversity. For a fixed degree sequence, the statistics of the effective competition matrix can only be modified by redistributing the pairwise overlaps. Such rearrangements may be designed to reduce $X$, but this optimization does not require the total nestedness $F$ to vary monotonically in either direction. In fact, we can introduce degree correlations in an HDMFT analysis,which  results in the mutualistic terms having additional noise and feedback terms proportional to the excess overlaps $f_{ik,jk}=P(ik,jk)-P(ik)P(jk)$, where $P(ij,kl)$ is the joint probability of two edges $i-j$ and $k-l$ being present. By considering the effect of nestedness on the mutualism for plant species, we obtain:  \begin{align}
    \gamma_{{\rm eff}(v)}^{'(P)}(t)=\gamma_{{\rm eff}(v)}^{(P)}(t)+\xi^{(2)}_v(t)+\int dt'du\rho(u)R^{(2)}_{vu}(t,t')A_u(t').\label{eq:DFMT_NODF}
\end{align} The additional noise here is also colored, with correlations  $\langle \xi^{(2)}_v(t)\xi^{(2)}_u(t')\rangle$ given by
\begin{align}
   \frac{\mu^2_\gamma}{c^2N}\langle  f_{wu,wv}P_w(t)P_w(t')\rangle_w+\frac{\mu^2_\gamma}c\langle  f_{uw,us}P_w(t)P_s(t')\rangle_{ws}\delta_{uv},
\end{align} with the response also depending on $\mu_\gamma$ and $f$ in similar ways. Doing a perturbational analysis of the fixed point with these additional degree correlations shows that the eigenvalues of the stability matrix are, on average, perturbed by the average of these excess overlaps, proportional to the sum $\sum_{ijk}f(ik,jk)$. This is the excess part of the same object $X$ defined above, i.e. how much higher its nestedness is than what is expected of a network with its given degree distribution. We conclude again that while a carefully constructed network with positive excess nestedness could possibly lead to higher stability, it normally won't. Note that this conclusion is heavily dependent on the definition of nestedness as the average of \textit{fractional} overlaps -- higher \textit{total} overlaps will always lead to less stable systems.

\onecolumngrid

\section{Appendix}

\subsection{Dynamical Mean-Field Theory in the case of a generic Monod function}

We aim to develop a dynamical treatment for an interacting model of $P$ plant and $A$ animal species whose mutual regulation occurs via a highly non-linear function.
We start from the dynamics presented in Saavedra, Stouffer, Bascompte \emph{et al.} Letter, in which the change in the abundance over time for a given plant of species $i$ is defined as
\begin{equation}
    \frac{d S_i^{(P)}}{dt}=\alpha_i^{(P)}S_i^{(P)}-\chi (S_i^{(P)})^2 -\sum_{j \in P} \beta_{ij}^{(P)} S_i^{(P)}S_j^{(P)} + \sum_{k \in A} \frac{\gamma_{ik}^{(P)}S_i^{(P)}S_k^{(A)}}{1+h^{(P)}\sum \limits_{l \in A}\gamma_{il}^{(P)}S_l^{(A)}}
\end{equation}
The analogous equation for animals is obtained by interchanging the superscripts $(P)$ and $(A)$. $\alpha_i$ denotes the growth rate that, for simplicity, we shall assume to be constant and species-independent. The matrices $\beta_{ij}$ and $\gamma_{ik}$ denote competitive and mutualistic couplings, respectively. We will consider both coefficients to be Gaussian distributed to perform the computations analytically. 
We shall take advantage of the usual path integral approach by introducing a fictitious source term that will be eventually sent to zero.

\begin{align}
\begin{aligned}
Z[\boldsymbol{\psi},\boldsymbol{\phi}]=\int & \mathcal{D}[P, \hat{P}] \exp \Biggl \lbrace i \sum_i \int dt  \hat{P}_i(t) \left[ \frac{\dot{P}_i}{P_i}-\left(\alpha -\chi P_i-\sum_{j } \beta^{(P)}_{ij} P_j+\sum \limits_{k } \frac{\gamma_{ik}^{(P)}A_k}{1+h^{(P)}\sum \limits_{l}\gamma_{il}^{(P)}A_l}  +\epsilon_i(t) \right) \right] \Biggr \rbrace  \\&\times \mathcal{D}[A, \hat{A}] \exp \Biggl \lbrace i \sum_i \int dt  \hat{A}_i(t) \left[ \frac{\dot{A}_i}{A_i}-\left(\alpha -\chi A_i-\sum_{j } \beta^{(A)}_{ij} A_j+\sum \limits_{k } \frac{\gamma_{ik}^{(A)}P_k}{1+h^{(A)}\sum \limits_{l }\gamma_{il}^{(A)}P_l}  +\xi_i(t) \right) \right] \Biggr \rbrace  \\&\times  \exp \left(i \sum_i \int dt \; P_i(t) \psi_i(t) \right) \times  \exp \left(i \sum_i \int dt \; A_i(t) \phi_i(t) \right) 
\end{aligned}
\end{align}
where we drop the superscripts for plants and animals instead writing them as $A$ and $P$ for ease. We assume the elements of the interaction matrices to be random and normally distributed, with mean and variance respectively:
\begin{align}
\begin{aligned}
    \beta^{(P)}_{ij}=\frac{\mu_P}{N_P}+\frac{\sigma_P}{\sqrt{N_P}} \eta_{ij}^{(P)} \hspace{1cm} \text{and} \hspace{1cm}
    \gamma^{(P)}_{ij}=\frac{m_P}{N_A}+\frac{s_P}{\sqrt{N_A}} z_{ij} ^{(P)},
\end{aligned}
\end{align}where the variables $\eta$ and $z$ have zero mean and variance 1. Additionally, the various sectors of interaction coefficients have the following correlations $\langle \eta_{ij}^{(A)} \eta_{ji}^{(A)} \rangle=c_A$, $\langle \eta_{ij}^{(P)} \eta_{ji}^{(P)} \rangle=c_P$, and $\langle z_{ij}^{(A)} z_{ji}^{(P)} \rangle=c_\gamma$. Performing the averages over the quenched disorder $\gamma_P=\frac{m_P}{N}+\frac{s_P}{\sqrt{N}}z^{(P)}$, we have
\begin{align}
\begin{aligned}
\int \prod_{\{ik\}} d z_{ik}^{(P)}d z_{ki}^{(A)}&\exp{\Biggr\lbrace-\frac{(z_{ki}^{(P)})^2+(z_{ik}^{(A)})^2-2c_\gamma z_{ik}^{(P)}z_{ki}^{(A)}}{1-c_\gamma^2}  -i\sum_{\{ik\}}  \int dt\left(\hat{a}_i (\frac{m_P}{N}+\frac{s_P}{\sqrt{N}}z_{ik}^{(P)}){A_k} +\hat{p}_k (\frac{m_A}{N}+\frac{s_A}{\sqrt{N}}z_{ki}^{(A)}){P_i} \right)\Biggr\rbrace}\\=\int\mathcal{D}[\dots]&\exp\left(-\frac{s_P^2}{2N}\sum_{ik}\int dt\int dt'\hat{a}_i(t)\hat{a}_i(t')A_k(t)A_k(t')-i\frac{m_P}{N}\sum_{ik}\int dt\hat{a}_i(t)A_k(t)\right)\\\times\int\mathcal{D}[\dots]&\exp\left(-\frac{s_A^2}{2N}\sum_{ik}\int dt\int dt'\hat{p}_i(t)\hat{p}_i(t')P_k(t)P_k(t')-i\frac{m_A}{N}\sum_{ik}\int dt\hat{p}_i(t)P_k(t)\right)\\\times&\exp\left(\frac{s_As_Pc_\gamma}{N}\sum_{ik}\int dt\int dt'\hat{a}_i(t)\hat{p}_k(t')A_k(t)P_i(t')\right)
\end{aligned}
\end{align}
Similarly taking the average over the $\beta$'s would give for example just for animals species
\begin{align}
    \begin{aligned}
        \prod_{ij}&\exp\left(i\int dt\hat{A}_i(t)A_j(t)\frac{\mu_A}{N}+\int\int dtdt'\hat{A}_i(t)A_j(t)\hat{A}_i(t')A_j(t')\frac{\sigma^2_A}{2N}\right)\\&\times\exp\left(-\int\int dtdt'\hat{A}_i(t)A_j(t)\hat{A}_j(t')A_i(t')\frac{\sigma^2_Ac_A}{N}\right),
    \end{aligned}
\end{align}with identical expressions for plant species. We then define the macroscopic parameters
\begin{align}
    \begin{aligned}
        L_A(t,t')=\frac{1}{N}\sum_i\hat{A}_i(t)\hat{A}_i(t'),\qquad &M_A(t,t')=\frac{1}{N}\sum_iA_i(t)A_i(t')\\
        L_P(t,t')=\frac{1}{N}\sum_i\hat{P}_i(t)\hat{P}_i(t'),\qquad &M_P(t,t')=\frac{1}{N}\sum_iP_i(t)P_i(t')\\
        K_P(t,t')=\frac{1}{N}\sum_iP_i(t)\hat{P}_i(t'),\qquad &K_A(t,t')=\frac{1}{N}\sum_iA_i(t)\hat{A}_i(t')\\
        \lambda_A(t)=\frac{1}{N}\sum_i\hat{A}_i(t),\qquad &\rho_A(t)=\frac{1}{N}\sum_iA_i(t)\\\lambda_P(t)=\frac{1}{N}\sum_i\hat{P}_i(t),\qquad &\rho_P(t)=\frac{1}{N}\sum_iP_i(t)\\
        r_a(t)=\frac{1}{N}\sum_i\hat{a}_i(t),\qquad &r_p(t)=\frac{1}{N}\sum_i\hat{p}_i(t)\\ 
        w_a(t,t')=\frac{1}{N}\sum_i\hat{a}_i(t)\hat{a}_i(t'),\qquad &w_p(t,t')=\frac{1}{N}\sum_i\hat{p}_i(t)\hat{p}_i(t')\\ 
        q_a(t)=\frac{1}{N}\sum_i\hat{a}_i(t)a_i(t),\qquad &q_p(t)=\frac{1}{N}\sum_i\hat{p}_i(t)p_i(t)\\ 
        v_a(t,t')=\frac{1}{N}\sum_i\hat{a}_i(t)P_i(t'),\qquad &v_p(t,t')=\frac{1}{N}\sum_i\hat{p}_i(t)A_i(t').
    \end{aligned}
\end{align}
This requires us to include the integral representations of these definitions, which is
\begin{align}
    \begin{aligned}
        \int \mathcal{D}[\dots]&\exp\left(iN\int dt\int dt'(\hat{L}_A(t,t')L_A(t,t')+\hat{M}_A(t,t')M_A(t,t'))+iN\int dt(\hat{\lambda}_A(t)\lambda_A(t)+\hat{\rho}_A(t)\rho_A(t))\right)\\\times&\exp\left(iN\int dt\int dt'(\hat{L}_P(t,t')L_P(t,t')+\hat{M}_P(t,t')M_P(t,t'))+iN\int dt(\hat{\lambda}_P(t)\lambda_P(t)+\hat{\rho}_P(t)\rho_P(t))\right)\\\times&\exp\left(iN\int dt\int dt'(\hat{K}_P(t,t')K_P(t,t')+\hat{K}_A(t,t')K_A(t,t'))+iN\int dt\int dt'(\hat{v}_a(t,t')v_a(t,t')+\hat{v}_p(t,t')v_p(t,t'))\right)\\\times&\exp\left(iN\int dt\int dt'(\hat{w}_a(t,t')w_a(t,t')+\hat{w}_p(t,t')w_p(t,t'))+iN\int dt(\hat{r}_a(t)r_a(t)+\hat{r}_p(t)r_p(t))\right)\\\times&\exp\left(iN\int dt(\hat{q}_a(t)q_a(t)+\hat{q}_p(t)q_p(t))\right)\\\times&\exp\left(-i\int dt\int dt'(\hat{L}_A\sum_i\hat{A}_i(t)\hat{A}_i(t')+\hat{M}_A\sum_iA_i(t)A_i(t'))-i\int dt(\hat{\lambda}_A\sum_i\hat{A}_i(t)+\hat{\rho}_A\sum_i A_i(t))\right)\\\times&\exp\left(-i\int dt\int dt'(\hat{K}_A\sum_i\hat{A}_i(t)A_i(t')+\hat{K}_P\sum_i\hat{P}_i(t)P_i(t'))-i\int dt\int dt'(\hat{v}_a\sum_i\hat{a}_i(t)P_i(t')+\hat{v}_p\sum_i\hat{p}_i(t)A_i(t'))\right)\\\times&\exp\left(-i\int dt\int dt'(\hat{L}_P\sum_i\hat{P}_i(t)\hat{P}_i(t')+\hat{M}_P\sum_iP_i(t)P_i(t'))-i\int dt(\hat{\lambda}_P\sum_i\hat{P}_i(t)+\hat{\rho}_P\sum_i P_i(t))\right)\\\times&\exp\left(-i\int dt\int dt'(\hat{w}_a\sum_i\hat{a}_i(t)\hat{a}_i(t')+\hat{w}_p\sum_i\hat{p}_i(t)\hat{p}_i(t'))-i\int dt(\hat{r}_a\sum_i\hat{a}_i(t)+\hat{r}_p\sum_i \hat{p}_i(t))\right)\\
        \times&\exp\left(-i\int dt(\hat{q}_a\sum_i\hat{a}_i(t)a_i(t)+\hat{q}_p\sum_i\hat{p}_i(t)p_i(t))\right)
    \end{aligned}
\end{align}
Here we'll call the first five lines $\exp N\Psi$ for ease. This lets us write the generating function as
\begin{align}
    \begin{aligned}
        \mathcal{Z}[\psi,\phi]=\int&\mathcal{D}[\dots]\exp(N\Psi)\\\times&\exp\left(-i\int dt\int dt'(\hat{L}_A\sum_i\hat{A}_i(t)\hat{A}_i(t')+\hat{M}_A\sum_iA_i(t)A_i(t'))-i\int dt(\hat{\lambda}_A\sum_i\hat{A}_i(t)+\hat{\rho}_A\sum_i A_i(t))\right)\\\times&\exp\left(-i\int dt\int dt'(\hat{L}_P\sum_i\hat{P}_i(t)\hat{P}_i(t')+\hat{M}_P\sum_iP_i(t)P_i(t'))-i\int dt(\hat{\lambda}_P\sum_i\hat{P}_i(t)+\hat{\rho}_P\sum_i P_i(t))\right)\\\times&\exp\left(-i\int dt\int dt'(\hat{w}_a\sum_i\hat{a}_i(t)\hat{a}_i(t')+\hat{w}_p\sum_i\hat{p}_i(t)\hat{p}_i(t'))-i\int dt(\hat{r}_a\sum_i\hat{a}_i(t)+\hat{r}_p\sum_i \hat{p}_i(t))\right)\\\times&\exp\left(-i\int dt\int dt'(\hat{K}_A\sum_i\hat{A}_i(t)A_i(t')+\hat{K}_P\sum_i\hat{P}_i(t)P_i(t'))-i\int dt\int dt'(\hat{v}_a\sum_i\hat{a}_i(t)P_i(t')+\hat{v}_p\sum_i\hat{p}_i(t)A_i(t'))\right)\\\times
& \exp \Biggl \lbrace i \sum_i \int dt  \hat{P}_i(t) \left[ \frac{\dot{P}_i}{P_i}-\left(\alpha -\chi P_i+\frac{a_i}{1+h^{(P)} a_i}  +\epsilon_i \right) \right] \Biggr \rbrace  \\\times
& \exp \Biggl \lbrace i \sum_i \int dt  \hat{A}_i(t) \left[ \frac{\dot{A}_i}{A_i}-\left(\alpha -\chi A_i+\frac{p_i}{1+h^{(A)} p_i}  +\xi_i \right) \right] \Biggr \rbrace \\\times&\exp\left(-\frac{Ns_P^2}{2}\int dt\int dt'w_a(t,t')M_A(t,t')-iNm_P\int dt r_a(t)\rho_A(t)+Ns_As_Pc_\gamma\int dt\int dt'v_a(t,t')v_p(t',t)\right)\\\times&\exp\left(-\frac{Ns_A^2}{2}\int dt\int dt'w_p(t,t')M_P(t,t')-iNm_A\int dt r_p(t)\rho_P(t)\right)\\\times&\exp\left(\frac{N\sigma_P^2}{2}\int dt\int dt'L_P(t,t')M_P(t,t')+iN\mu_P\int dt \lambda_P(t)\rho_P(t)-N\sigma_P^2c_P\int dt\int dt'K_P(t,t')K_P(t',t)\right)\\\times&\exp\left(\frac{N\sigma_A^2}{2}\int dt\int dt'L_A(t,t')M_A(t,t')+iN\mu_A\int dt \lambda_A(t)\rho_A(t)-N\sigma_A^2c_A\int dt\int dt'K_A(t,t')K_A(t',t)\right)\\
\times&\exp\left(iN\int dt q_a(t)+iN\int dt q_p(t)\right)\exp\left(-i\int dt(\hat{q}_a\sum_i\hat{a}_i(t)a_i(t)+\hat{q}_p\sum_i\hat{p}_i(t)p_i(t))\right)\\ \times&\exp \left(i \sum_i \int dt \; P_i(t) \psi_i(t) \right) \exp \left(i \sum_i \int dt \; A_i(t) \phi_i(t) \right) 
    \end{aligned}
\end{align}
Which at the saddle point should give via maximisation in hatted variables
\begin{align}
    \begin{aligned}
        L_A=\langle \hat{A}\hat{A}'\rangle&\qquad M_A=\langle AA'\rangle\\
        L_P=\langle \hat{P}\hat{P}'\rangle&\qquad M_P=\langle PP'\rangle\\
        K_P=\langle \hat{P}P'\rangle&\qquad K_A=\langle \hat{A}A'\rangle\\
\lambda_A=\langle \hat{A}\rangle&\qquad \rho_A=\langle A\rangle\\
\lambda_P=\langle \hat{P}\rangle&\qquad \rho_P=\langle P\rangle\\
w_a=\langle \hat{a}\hat{a}'\rangle&\qquad w_p=\langle \hat{p}\hat{p}'\rangle\\
r_a=\langle \hat{a}\rangle&\qquad p_a=\langle \hat{p}\rangle\\
q_a=\langle \hat{a}a\rangle&\qquad q_p=\langle \hat{p}p\rangle\\
v_a=\langle \hat{a}A\rangle&\qquad v_p=\langle \hat{p}P\rangle
\end{aligned}
\end{align}
which are just basically the definitions of the variables. Maximising over the regular variables now we have
\begin{align}
    \begin{aligned}
         i\hat{L}_A=-\frac{1}{2}\sigma_A^2M_A&\qquad i\hat{M}_A=-\frac{1}{2}\sigma_A^2L_A+\frac1{2}s_P^2w_a\\
         i\hat{L}_P=-\frac{1}{2}\sigma_P^2M_P&\qquad i\hat{M}_P=-\frac{1}{2}\sigma_P^2L_P+\frac12s_A^2w_p\\
         i\hat{K}_P(t,t')=c_P\sigma_P^2K_P(t',t)&\qquad i\hat{K}_A(t,t')=c_A\sigma_A^2K_A(t',t)\\
         \hat{\lambda}_P=-\mu_P\rho_P&\qquad \hat{\rho}_P=-\mu_P\lambda_P+m_Ar_p\\
         \hat{\lambda}_A=-\mu_A\rho_A&\qquad \hat{\rho}_A=-\mu_A\lambda_A+m_Pr_a\\
         i\hat{w}_a=\frac1{2}s_P^2M_A&\qquad \hat{r}_a=m_P\rho_A\\
         i\hat{w}_p=\frac1{2}s_A^2M_P&\qquad \hat{r}_p=m_A\rho_P\\
         \hat{q}_a=-1&\qquad\hat{q}_p=-1\\
         i\hat{v}_a(t,t')=-s_Ps_Ac_\gamma v_p(t',t)&\qquad i\hat{v}_p(t,t')=-s_Ps_Ac_\gamma  v_a(t',t)
    \end{aligned}
\end{align}
As the average over hatted $A$'s and $P$'s must be zero, $L_A=L_P=\lambda_A=\lambda_P=0$. A similar line of reasoning should also be applying to hatted $a$ and $p$, giving $r_a=p_a=w_a=w_p=0$, and $\hat{M}_A=\hat{M}_P=\hat{\rho}_A=\hat{\rho}_P=0$. The entire generating function then becomes
\begin{align}
    \begin{aligned}
        \mathcal{Z}[\psi,\phi]=\int\mathcal{D}[\dots]\times&\exp\left(iN\int dt\int dt'\hat{L}_A\hat{A}(t)\hat{A}(t')+iN\int dt\hat{\lambda}_A\hat{A}(t)\right)\\\times&\exp\left(iN\int dt\int dt'\hat{L}_P\hat{P}(t)\hat{P}(t')+iN\int dt\hat{\lambda}_P\hat{P}(t)\right)\\\times&\exp\left(-iN\int dt\int dt'(\hat{w}_a\hat{a}(t)\hat{a}(t')+\hat{w}_p\hat{p}(t)\hat{p}(t'))-iN\int dt(\hat{r}_a\hat{a}(t)+\hat{r}_p\hat{p}(t))\right)\\\times&\exp\left(-i\int dt\int dt'(\hat{K}_A\hat{A}(t)A(t')+\hat{K}_P\hat{P}(t)P(t'))\right)\\\times&\exp\left(-i\int dt\int dt'(\hat{v}_a\hat{a}(t)P(t')+\hat{v}_p\hat{p}(t)A(t'))\right)\\\times
& \exp \Biggl \lbrace i N \int dt  \hat{P}(t) \left[ \frac{\dot{P}}{P}-\left(\alpha -\chi P+\frac{a}{1+h^{(P)} a}  +\epsilon \right) \right] \Biggr \rbrace  \\\times
& \exp \Biggl \lbrace i N\int dt  \hat{A}(t) \left[ \frac{\dot{A}}{A}-\left(\alpha -\chi A+\frac{p}{1+h^{(A)} p}  +\xi \right) \right] \Biggr \rbrace \\\times&\exp\left(-iN\int dt(\hat{q}_a\hat{a}(t)a(t)+\hat{q}_p\hat{p}(t)p(t))\right)\\ \times&\exp \left(i N\int dt \; P(t) \psi(t) \right) \exp \left(i N\int dt \; A(t) \phi(t) \right) 
    \end{aligned}
\end{align}
which helps us separate out the terms in each variable, getting
\begin{itemize}
    \item $a(t)$: \begin{align}
        \exp\left[iN\int dt \hat{a}(t)\left(a(t)-m_P\rho_A(t)-s_Ps_Ac_\gamma \int dt'v_p(t',t)P(t')-\eta_a(t)\right)\right]
    \end{align}
    where $\langle \eta_a(t)\eta_a(t')\rangle=s_P^2M_A(t,t')$. This gives $a(t)=m_P\rho_A(t)+\int dt' s_Ps_Ac_\gamma v_p(t',t)P(t')+\eta_a(t)$.
    \item $p(t)$: \begin{align}
        \exp\left[iN\int dt \hat{p}(t)\left(p(t)-m_A\rho_P(t)-s_As_Pc_\gamma \int dt'v_a(t',t)A(t')-\eta_p(t)\right)\right]
    \end{align}
    where $\langle \eta_p(t)\eta_p(t')\rangle=s_A^2M_P(t,t')$. This gives $p(t)=m_A\rho_P(t)+\int dt' s_As_Pc_\gamma v_a(t',t)A(t')+\eta_p(t)$.
    \item $A(t)$: \begin{align}
        \exp\left[iN\int dt \hat{A}(t)\left(\frac{\dot{A}}{A}-\alpha+\chi A-\frac{p}{1+h^{(A)}p}+\mu_A\rho_A(t)+c_A\sigma^2_A\int dt' K_A(t',t)A(t')-\eta_A+\epsilon\right)\right]
    \end{align}
    where  $\langle \eta_A(t)\eta_A(t')\rangle=\sigma_A^2M_A(t,t').$
    \item $P(t)$: \begin{align}
        \exp\left[iN\int dt \hat{P}(t)\left(\frac{\dot{P}}{P}-\alpha+\chi P-\frac{a}{1+h^{(P)}a}+\mu_P\rho_P(t)+c_P\sigma^2_P\int dt' K_P(t',t)P(t')-\eta_P+\xi\right)\right]
    \end{align}
    where  $\langle \eta_P(t)\eta_P(t')\rangle=\sigma_P^2M_P(t,t').$
\end{itemize}
This gives us the following final equations in species populations
\begin{align}
    \begin{aligned}
        \frac{\dot{P}(t)}{P(t)}&=\alpha-\chi P(t)+\frac{\eta_a(t)+m_P\rho_A(t)+\int dt' c_\gamma s_Ps_A v_p(t',t)P(t')}{1+h^{(P)}(\eta_a(t)+m_P\rho_A(t)+\int dt' c_\gamma s_Ps_A v_p(t',t)P(t'))}-\mu_P\rho_P(t)+\eta_P(t)+\int dt' c_P \sigma_P^2 K_P(t',t)P(t')\\
        \frac{\dot{A}(t)}{A(t)}&=\alpha-\chi A(t)+\frac{\eta_p(t)+m_A\rho_P(t)+\int dt' c_\gamma s_As_P v_a(t',t)A(t')}{1+h^{(A)}(\eta_p(t)+m_A\rho_P(t)+\int dt' c_\gamma s_As_P v_a(t',t)A(t'))}-\mu_A\rho_A(t)+\eta_A(t)+\int dt' c_A \sigma_A^2 K_A(t',t)A(t')
    \end{aligned}
\end{align}

\subsection{Fixed point analysis}

Doing a fixed point analysis in the small $h$ limit, we have
\begin{align}
    \begin{aligned}
        P^*\left(\alpha-\chi P^*+\frac{\sqrt{M_A^*}s_Pz_a+m_P\rho_A^*}{1+h ^{(P)}(\sqrt{M_A^*}s_Pz_a+m_P\rho_A^*)}-\mu_P\rho_P^*-\sqrt{M_P^*}\sigma_Pz_P\right)&=0\\
        A^*\left(\alpha-\chi A^*+\frac{\sqrt{M_P^*}s_Az_p+m_A\rho_P^*}{1+h^{(A)}(\sqrt{M_P^*}s_Az_p+m_A\rho_P^*)}-\mu_A\rho_A^*-\sqrt{M_A^*}\sigma_Az_A\right)&=0\\\implies \chi P^*={\rm max}\left(0,\alpha+(\sqrt{M_A^*}s_Pz_a+m_P\rho_A^*)-h^{(P)}(\sqrt{M_A^*}s_Pz_a+m_P\rho_A^*)^2-\mu_P\rho_P^*-\sqrt{M_P^*}\sigma_Pz_P\right)&\\
        \implies \chi A^*={\rm max}\left(0,\alpha+(\sqrt{M_P^*}s_Az_p+m_A\rho_P^*)-h^{(A)}(\sqrt{M_P^*}s_Az_p+m_A\rho_P^*)^2-\mu_A\rho_A^*-\sqrt{M_A^*}\sigma_Az_A\right)&
    \end{aligned}
\end{align}
For self consistency we need
\begin{align}
    \begin{aligned}
        \rho_A^*=\langle A^*\rangle_{\{z\}}&\qquad\rho_P^*=\langle P^*\rangle_{\{z\}}\\
        M_A^*=\langle A^*A^*\rangle_{\{z\}}&\qquad M_P^*=\langle P^*P^*\rangle_{\{z\}}
    \end{aligned}
\end{align}
where we know all the $z$'s have zero mean and unit variance, and are assumed to be independent of each other. Attempting to simplify the expressions above,
\begin{align}
    \begin{aligned}
        \chi A^*&=\left(a+bz_1+c-dz_2-h(a+bz_1)^2\right)_+=\left(a(1-ha)-c+b(1-2ah)z_1-hb^2z_1^2-dz_2\right)_+        \\
        &\approx \left(a(1-ha)-c+b(1-2ha)z_1'-hb^2-dz_2\right)_+\\
        &=\left(a(1-ha)-c-hb^2+Z\sqrt{b^2(1-2ah)^2+d^2}\right)_+\\
        \rho_A^*&=\int_{-\frac{a(1-ha)-c-hb^2}{\sqrt{b^2(1-2ah)^2+d^2}}}^\infty dZ\left(a(1-ha)-c-hb^2+Z\sqrt{b^2(1-2ah)^2+d^2}\right)e^{-Z^2/2}\\
        M_A^*&=\int_{-\frac{a(1-ha)+c-hb^2}{\sqrt{b^2(1-2ah)^2+d^2}}}^\infty dZ\left(a(1-ha)-c-hb^2+Z\sqrt{b^2(1-2ah)^2+d^2}\right)^2e^{-Z^2/2}\\\phi&=\int_{-\frac{a(1-ha)-c-hb^2}{\sqrt{b^2(1-2ah)^2+d^2}}}^\infty dZe^{-Z^2/2}
    \end{aligned}
\end{align}
The dynamical dmft equation in these terms is
\begin{align}
    \begin{aligned}
        \dot{S}&=S^*(a+bz_1-S^*-h(a+bz_1)^2-c-dz_2+\eta)\\
        \delta\dot{S}&=S^*(b\delta z_1-\delta S-2hab\delta z_1-d\delta z_2+\delta\eta)\\
        i\omega\delta\tilde{S}&=S^*(b\delta \tilde{z}_1-\delta \tilde{S}-2hab\delta \tilde{z}_1-d\delta \tilde{z}_2+\delta\tilde\eta)\\
        \delta \tilde{S}&=\frac{b\delta \tilde{z}_1-2hab\delta \tilde{z}_1-d\delta \tilde{z}_2+\delta\tilde\eta}{i\omega/S^*+1}\\
        \langle|\delta\tilde{S}|\rangle^2&=\phi b^2(1-4ha)+\phi d^2+1\\
        \langle|\delta\tilde{S}|\rangle^2&=\frac{1}{1-\phi s^2(1-4h\rho m)-\phi \sigma^2}\\
        \implies 1&=\phi_cs_c^2(1-4h\rho m)+\phi_c\sigma_c^2
    \end{aligned}
\end{align}
rewriting the self consistency equations
\begin{align}
    \begin{aligned}
        \rho^*&=B\int_{-\Delta}^\infty DZ(-\Delta+Z)=\frac{B}{2}\left[e^{-\Delta^2/2}\sqrt{\frac2\pi}+\Delta(1+{\rm erf}\frac{\Delta}{\sqrt{2}} )\right]\\
        M^*&=B^2\int_{-\Delta}^\infty DZ(-\Delta+Z)^2=\frac{B^2}{2}(1+\Delta^2)(1+{\rm erf} \frac{\Delta}{\sqrt{2}})+\frac{B^2}{\sqrt{2\pi}}e^{-\Delta^2/2}\Delta\\
        \phi&=\int_{-\Delta}^\infty DZ\equiv\frac{1}{2}(1+{\rm erf}\frac{\Delta}{\sqrt{2}})=_cM^*/B^2
    \end{aligned}
\end{align}
These along with the criticality condition above condition above gives $\phi=1/2, \Delta=0$, and
\begin{align}
    \begin{aligned}
        \rho^*&=\sqrt{\frac{Ms^2(1-4\rho h m)+M\sigma^2}{2\pi}}\\
        0&=1-hMs^2+m\rho(1-hm\rho)-\mu\rho\\
        2&=s^2(1-4hm\rho)+\sigma^2
    \end{aligned}
\end{align}
which gives to first order in $h$
\begin{align}
    \begin{aligned}
        \rho^*=\frac{1}{\mu-m}+h\frac{\pi s^2+m^2}{(\mu-m)^3},
    \end{aligned}
\end{align}
and thus for criticality,
\begin{align}
    \begin{aligned}
        s_c^2\left(1-\frac{4hm}{\mu-m}\right)+\sigma_c^2=2
    \end{aligned}
\end{align}
In these approximations the stationary distribution is also just a truncated gaussian.

\subsection{Linear stability analysis}
Consider again the dynamical equations 
\begin{align}
    \frac{d S_i^{(P)}}{dt}=S_i^{(P)}\left(1- S_i^{(P)} -\sum_{j \in P} \beta_{ij}^{(P)} S_j^{(P)} + \sum_{k \in A} \frac{\gamma_{ik}^{(P)}S_k^{(A)}}{1+h^{(P)}\sum \limits_{l \in A}\gamma_{il}^{(P)}S_l^{(A)}}\right).
\end{align}

Around a supposed fixed point given by the equilibrium abundances $\{S_i^{(P)*}\}$ and $\{S_i^{(A)*}\}$, we can perform a linear perturbation to get

\begin{align}
    \begin{aligned}
        \begin{pmatrix}
            \Dot{\delta \Vec{S}}^{(P)}\\
            \Dot{\delta \Vec{S}}^{(A)}
        \end{pmatrix}_{2N\times 1}=\begin{pmatrix}
            \mathbb{A}& \mathbb{B}\\\mathbb{C} &\mathbb{D}
        \end{pmatrix}_{2N\times 2N}\begin{pmatrix}
            \delta \Vec{S}^{(P)}\\
            \delta \Vec{S}^{(A)}
        \end{pmatrix}_{2N\times 1}
    \end{aligned}
\end{align}
where the stability matrix is written as a $2\times2$ matrix comprised of blocks of size $N\times N$. We find that the full jacobian has blocks with the expressions
\begin{align}
    \begin{aligned}
        \mathbb{A}_{ij}&=\delta_{ij}\left(1-2S_i^{(P)*}-\sum_k\beta_{ik}^{(P)}S_k^{(P)*}+\frac{\sum_k\gamma_{ik}^{(P)}S_k^{(A)*}}{1+h^{(P)}\sum_l\gamma_{il}^{(P)}S_l^{(A)*}}\right)-S_i^{(P)*}\beta_{ij}^{(P)}\\
        \mathbb{B}_{ij}&=\frac{\gamma_{ij}^{(P)}S_i^{(P)*}}{(1+h^{(P)}\sum_l\gamma_{il}^{(P)}S_l^{(A)*})^2}\\
        \mathbb{C}_{ij}&=\frac{\gamma_{ij}^{(A)}S_i^{(A)*}}{(1+h^{(A)}\sum_l\gamma_{il}^{(A)}S_l^{(P)*})^2}\\
        \mathbb{D}_{ij}&=\delta_{ij}\left(1-2S_i^{(A)*}-\sum_k\beta_{ik}^{(A)}S_k^{(A)*}+\frac{\sum_k\gamma_{ik}^{(A)}S_k^{(P)*}}{1+h^{(A)}\sum_l\gamma_{il}^{(A)}S_l^{(P)*}}\right)-S_i^{(A)*}\beta_{ij}^{(A)}
    \end{aligned}
\end{align}
In the single equilibrium limit without zero abundances, small values of $h$, and same statistics for plant and animals species, these can be simplified to write in the notation of \cite{patil2024spectral} to give the covariance matrices
\begin{align}
    \begin{aligned}
        \Psi&=\begin{pmatrix}
            \sigma_\beta^2&0&0&0\\
            0&\sigma_\gamma^2(1-4hS^*+4h^2S^{2*}\sigma_\gamma^2)&0&0\\
            0&0&\sigma_\gamma^2(1-4hS^*+4h^2S^{2*}\sigma_\gamma^2)&0\\
            0&0&0&\sigma_\beta^2
        \end{pmatrix},\\
        \Upsilon&=\begin{pmatrix}
            \sigma_\beta^2\rho_\beta&0&0&0\\
            0&0&\sigma_\gamma^2(1-4hS^*+4h^2S^{2*}\sigma_\gamma^2\rho_\gamma)\rho_\gamma&0\\
            0&\sigma_\gamma^2(1-4hS^*+4h^2S^{2*}\sigma_\gamma^2\rho_\gamma)\rho_\gamma&0&0\\
            0&0&0&\sigma_\beta^2\rho_\beta
        \end{pmatrix}.
    \end{aligned}
\end{align}
where the terms in $\Psi$ denote covariances of an element $ij$ of one block with the same of another, while $\Upsilon$ contains the covariances of terms $ij$ of one block with $ji$ of another. These give the same coefficients to the various terms as the DMFT perturbation analysis, but lack information on the equilibrium value of $S^*$ needed to complete the calculations.

\subsection{Analysis of effective langevin}
We should be able to write a single effective langevin when plant and animal interactions statistics are equivalent as follows
\begin{align}
\begin{aligned}
   \frac{\dot{S}(t)}{S(t)}&=1-S(t)+\frac{\eta_s(t)+m\rho(t)+\int dt' c_\gamma s^2 v_s(t',t)S(t')}{1+h(\eta_s(t)+m\rho(t)+\int dt' c_\gamma s^2 v_s(t',t)S(t'))}-\mu\rho(t)+\eta_S(t)+\int dt' c_\beta \sigma^2 K(t',t)S(t')
\end{aligned}
\end{align}
which in equilibrium is
\begin{align}
    \begin{aligned}
        0&=S^*\left(1-S^*+\frac{\eta_s+mM^*+c_\gamma s^2 \chi_sS^*}{1+h(\eta_s+mM^*+c_\gamma s^2 \chi_sS^*)}-\mu M^*+\eta_S+c_\beta \sigma^2 \chi_SS^*\right)
    \end{aligned}
\end{align}
which gives $S^*=f(\xi,\zeta)\Theta(f(\xi,\zeta))$, 
where
\begin{align}
\begin{aligned}
    f&=\frac{y-x+h\zeta y-h\xi x+\sqrt{(y-x+hy\zeta-hx\xi)^2+4yx(\xi+\zeta+h\xi\zeta)}}{2hyx}\\
    &=\frac{2(\xi+\zeta+h\xi\zeta)}{\sqrt{(y-x+hy\zeta-hx\xi)^2+4yx(\xi+\zeta+h\xi\zeta)}-y+x-h\zeta y+h\xi x}\\
    y&=c_\gamma s^2\chi_s\\
    x&=1-c_\beta \sigma^2\chi_S\\
    \xi&=\eta_s+mM\\
    \zeta&=1-\mu M+\eta_S
\end{aligned}
\end{align}
and $\eta_s=s\sqrt{q}z$, $\eta_S=\sigma\sqrt{q}Z$. For self consistency we would have
\begin{align}
    \begin{aligned}
        M&=\langle S^*\rangle=\int DzDZf(\xi,\zeta)\Theta(\xi+\zeta+h\xi\zeta)\\
        q&=\langle S^{2*}\rangle=\int DzDZf(\xi,\zeta)^2\Theta(\xi+\zeta+h\xi\zeta)\\
        \chi_s&=\Big\langle\frac{\delta S^*}{\delta \eta_s}\Big\rangle=\int DzDZ\frac{1}{s\sqrt{q}}\frac{\partial f(\xi,\zeta)\Theta(\xi+\zeta+h\xi\zeta)}{\partial z}\\
        \chi_S&=\Big\langle\frac{\delta  S^*}{\delta \eta_S}\Big\rangle=\int DzDZ\frac{1}{\sigma\sqrt{q}}\frac{\partial f(\xi,\zeta)\Theta(\xi+\zeta+h\xi\zeta)}{\partial Z}
    \end{aligned}
\end{align}
which for the response functions can be written as
\begin{align}
    \begin{aligned}
        s\sqrt{q}\chi_s&=\int dz \int dZ e^{-z^2/2}e^{-Z^2/2}\frac{\partial f(z,Z)\Theta(C+z)}{\partial z}\\
        &=\int dz \int dZ z  e^{-z^2/2}e^{-Z^2/2}f(z,Z)\Theta(C+z)\\
        \chi_s&=\frac{1}{s\sqrt{q}}\langle z S^* \rangle,\qquad \chi_S=\frac{1}{\sigma\sqrt{q}}\langle Z S^* \rangle.
    \end{aligned}
\end{align}
The species abundance distribution should be given by
\begin{align}
    \begin{aligned}
        P_{\rm st}(S^*)&=\langle\delta(S^*-f(\xi,\zeta)\Theta(\xi+\zeta+h\xi\zeta))\rangle_{\xi,\zeta}\\
        &=(1-\phi)\delta(S^*)+P_{\rm surv}(S^*)\Theta(S^*)
    \end{aligned}
\end{align}
The simplest form of the function $f$ we could work with is
\begin{align}
    \begin{aligned}
        f(z_1,z_2)&=1-\mu M+\sigma\sqrt{q}z_1+mM+s\sqrt{q}z_2-hm^2M^2-2hmMs\sqrt{q}z_2-hs^2qz_2^2\\
        f(Z,z)&=1-\mu M+mM-hm^2M^2-\sqrt{\sigma^2q+s^2q(1-2hmM)^2}Z-hs^2qz_2^2
    \end{aligned}
\end{align}
giving
\begin{align}
    \begin{aligned}
        P_{\rm st}&=\int DZ\int Dz\delta(S^*-f(Z,z)\Theta(f(Z,z)))\\
        &=\delta (S^*)\int_{-\infty}^\infty Dz\int^{\infty}_{\Delta(z^2)}DZ+\frac1{2\pi}\Theta(S^*)\int_{-\infty}^{\infty}\exp(-z^2/2)\exp\left((\Delta(z^2)-\frac{S^*}{\sqrt{\sigma^2q+s^2q(1-2hmM)^2}})^2\right)
    \end{aligned}
\end{align}
Ignoring the order four term in $z$ because of its $h^2$ coefficient in the exponent, we will have something of the sort\begin{align}
    P(S)=\frac{e^{-(\frac{S-a}{b})^2}}{\sqrt{c+hdS}}\Theta(S)
\end{align}

\subsection{Non-trivial degree distributions for mutualism}

Consider the same system as before, but without having each animal species affect all plant species and vice versa. The dynamical equations can be written in terms of the adjacency matrix $C_{ij}$ of such a network, 
\begin{align}
    \begin{aligned}
         \frac{d S_i^{(P)}}{dt}=S_i^{(P)}\left(1- S_i^{(P)} -\sum_{j \in P} \beta_{ij}^{(P)} S_j^{(P)} + \sum_{k \in A} \frac{C_{ik}\gamma_{ik}^{(P)}S_k^{(A)}}{1+h^{(P)}\sum \limits_{l \in A}C_{il}\gamma_{il}^{(P)}S_l^{(A)}}\right).
    \end{aligned}
\end{align}
This slightly modifies the generating function to give
\begin{align}
\begin{aligned}
Z[\boldsymbol{\psi},\boldsymbol{\phi}]=\int & \mathcal{D}[\dots] \int \prod_i  \frac{d a_i d\hat{a}_i}{2 \pi}\frac{d p_i d\hat{p}_i}{2 \pi}  \exp \Biggl \lbrace i \sum_i \int dt\hat{a}_i \left(a_i -\sum_k C_{ik}\gamma_{ik}^{(P)}A_k \right) \Biggr \rbrace 
\exp \Biggl \lbrace i \sum_i \int dt\hat{p}_i \left(p_i -\sum_kC_{ki} \gamma_{ik}^{(A)}P_k \right) \Biggr \rbrace \\\times
& \exp \Biggl \lbrace i \sum_i \int dt  \hat{P}_i(t) \left[ \frac{\dot{P}_i}{P_i}-\left(\alpha -\chi P_i-\sum_{j} \beta_{ij}^{(P)} P_j+\frac{a_i}{1+h^{(P)} a_i}  +\epsilon_i \right) \right] \Biggr \rbrace  \\\times
& \exp \Biggl \lbrace i \sum_i \int dt  \hat{A}_i(t) \left[ \frac{\dot{A}_i}{A_i}-\left(\alpha -\chi A_i-\sum_{j } \beta_{ij}^{(A)} A_j+\frac{p_i}{1+h^{(A)} p_i}  +\xi_i \right) \right] \Biggr \rbrace 
 \\& \times\exp \left(i \sum_i \int dt \; P_i(t) \psi_i(t) \right) \exp \left(i \sum_i \int dt \; A_i(t) \phi_i(t) \right) \prod_i\\
 =\int &\mathcal D[\dots]\mathcal M_A\mathcal M_P\exp \left(i \sum_i \int dt \; P_i(t) \psi_i(t) \right) \exp \left(i \sum_i \int dt \; A_i(t) \phi_i(t) \right)\\
 &\times \exp\Biggr\lbrace- i\sum_i\int dt\left[\hat P_i(t)\sum_j\beta_{ij}^{(P)}P_j+\hat A_i(t)\sum_j\beta_{ij}^{(A)}A_j+\hat a_i\sum_j C_{ij}\gamma_{ij}^{(P)}A_j+\hat p_i\sum_jC_{ji}\gamma_{ij}^{(A)}P_j\right]\Biggr\rbrace
\end{aligned}
\end{align}
where
\begin{align}
    \begin{aligned}
        \mathcal M_A&=\exp\Biggl \lbrace i \sum_i  \int dt  \left[ \hat{A}_i(t)\left( \frac{\dot{A}_i}{A_i}-1 + A_i-\frac{p_i}{1+h^{(A)} p_i} -\xi_i \right)  +\hat a_i(t)a_i(t)\right]\Biggr \rbrace \\
        \mathcal M_P&=\exp\Biggl \lbrace i \sum_i  \int dt  \left[ \hat{P}_i(t)\left( \frac{\dot{P}_i}{P_i}-1 + P_i-\frac{a_i}{1+h^{(P)} a_i} -\epsilon_i \right)  +\hat p_i(t)p_i(t)\right]\Biggr \rbrace 
    \end{aligned}
\end{align}
Now consider only the averages over $\gamma$ and $C$ using the methods of \cite{aguirre2024heterogeneous,park2024incorporating,poley2025interaction,patil2026dynamical}, 
\begin{align}
    \begin{aligned}
        &\overline{\exp\Biggr\lbrace-i\sum_{ij}C_{ij}\left(\hat a_i\gamma^{(P)}_{ij}A_j+\hat p _j\gamma^{(A)}_{ji}P_i\right)\Bigg\rbrace}\\
        =&\overline{\prod_{ij}\exp\Biggr\lbrace-iC_{ij}\left(\hat a_i\gamma^{(P)}_{ij}A_j+\hat p _j\gamma^{(A)}_{ji}P_i\right)\Bigg\rbrace}\\
        =&\overline{\prod_{ij}1+P(C_{ij}=1)(\exp\Biggr\lbrace-i\left(\hat a_i\gamma^{(P)}_{ij}A_j+\hat p _j\gamma^{(A)}_{ji}P_i\right)\Bigg\rbrace-1)}\\
        =&\overline{\prod_{ij}1+P(k_i\kappa_j)(\exp\Biggr\lbrace-i\left(\hat a_i\gamma^{(P)}_{ij}A_j+\hat p _j\gamma^{(A)}_{ji}P_i\right)\Bigg\rbrace-1)}\\
        =&\prod_{ij}\left(1+P(k_i\kappa_j)(\exp\Biggr\lbrace-i\left(\frac {\hat a_iA_jm_P}{C_A}+\frac {\hat p _jP_im_A}{C_P}\right)-\frac{\hat a_i\hat a_i'A_jA_j's_P^2}{2C_A}-\frac{\hat p_j\hat p_j'P_iP_i's_A^2}{2C_P}-\frac{\hat a_i\hat p_j'A_jP_i's_Ps_Ac_\gamma}{C}\Bigg\rbrace-1)\right)\\
        \approx&\prod_{ij}\exp\Biggr\lbrace\left[\sum_na_n^2\left({-k_i\kappa_j}\right)^n\right]\left[-i\left(\frac {\hat a_iA_jm_P}{C_A}+\frac {\hat p _jP_im_A}{C_P}\right)-\frac{\hat a_i\hat a_i'A_jA_j's_P^2}{2C_A}-\frac{\hat p_j\hat p_j'P_iP_i's_A^2}{2C_P}-\frac{\hat a_i\hat p_j'A_jP_i's_Ps_Ac_\gamma}{C}\right]\Biggr\rbrace
    \end{aligned}
\end{align}
Here $P(C_{ij}=1)=P(k_i\kappa_j)$ is the probability of an animal species with degrees $\kappa_j$ being connected to a plant species with $k_i$ degrees. The scalings of the interaction terms give the denominator $C_A$ (or $C_P$) which is the average number of animals (or plants) a plant (or animal) species interacts with, and $C=\sqrt{C_AC_P}$.  We define the order parameters
\begin{align}
    \begin{aligned}
        b_n(t)&=\frac{a_n}{N_P}\sum_i(-k_i)^n\hat a_i(t)&\qquad B_n&= \frac{a_n}{N_A}\sum_i(\kappa_i)^n A_i(t)\\
        d_n(t)&=\frac{a_n}{N_A}\sum_i(-\kappa_i)^n\hat p_i(t)&\qquad D_n&= \frac{a_n}{N_P}\sum_i(k_i)^nP_i(t)\\
        k_n(t,t')&=\frac{a_n}{N_P}\sum_i(-k_i)^n\hat a_i(t)\hat a_i(t')&\qquad K_n(t,t')&=\frac{a_n}{N_A}\sum_i(\kappa_i)^n A_i(t)A_i(t')\\
        g_n(t,t')&=\frac{a_n}{N_A}\sum_i(-\kappa_i)^n\hat p_i(t)\hat p_i(t')&\qquad G_n(t,t')&=\frac{a_n}{N_P}\sum_i(k_i)^n P_i(t)P_i(t')\\
        y_n(t,t')&=\frac{a_n}{N_A}\sum_i(-\kappa_i)^n\hat p_i(t)A_i(t')&\qquad z_n(t,t')&=\frac{a_n}{N_P}\sum_i(k_i)^n\hat a_i(t)P_i(t').
    \end{aligned}
\end{align} These are combinations of the terms obtained on averaging the interactions with various exponential orders of the degrees. Writing down the whole generating function again
\begin{align}
    \begin{aligned}
        \int &\mathcal D[\dots]\mathcal M_A\mathcal M_P\exp \left(i \sum_i \int dt \; P_i(t) \psi_i(t) \right) \exp \left(i \sum_i \int dt \; A_i(t) \phi_i(t) \right)\\
 &\times \exp\Biggr\lbrace- iN_A\int dt\left[\mu_A\lambda_A\rho_A+f_P\mu_P\lambda_P\rho_P+\int dt'\left(\frac12\sigma_A^2L_AM_A+f_P\frac12\sigma_P^2L_PM_P+\sigma_A^2c_AK_AK_A^T+f_P\sigma_P^2c_PK_PK_P^T\right)\right]\Biggr\rbrace\\&\times\exp\Biggr\lbrace-iN_A\sum_n\int dt\left[f_P\frac {m_P}{\langle v_A\rangle}b_nB_n+\frac {m_A}{\langle v_P\rangle}d_nD_n+\int dt'\left(\frac {s_A^2}{2\langle v_P\rangle}g_nG_n+f_P\frac {s_P^2}{2\langle v_A\rangle}k_nK_n+\sqrt{f_P}\frac {s_As_Pc_\gamma}{\langle v\rangle}y_nz_n\right)\right]\Biggr\rbrace\\
 &\times({\rm all\; parameter\;definitions})
    \end{aligned}
\end{align} we will find that $b,d,k,$ and $g$ are $0$, leading to $\hat B,\hat D,\hat K,$ and $\hat G$ being zero as well. The non-zero parameters will have the relations 
\begin{align}
    \begin{aligned}
        \hat b&=f_P\frac{m_P}{\langle v_A\rangle} B&\qquad \hat d&= \frac{m_A}{\langle v_P\rangle} D\\
        \hat k&=f_P\frac{s_P^2}{2{\langle v_A\rangle}}K&\qquad\hat g&=\frac{s_A^2}{2{\langle v_P\rangle}}G\\
        \hat y(t,t')&=\frac{\sqrt{f_P}}{\langle v\rangle}s_Ps_Ac_\gamma z(t',t)&\qquad\hat z(t,t')&=\frac{\sqrt{f_P}}{\langle v\rangle}s_Ps_Ac_\gamma y(t',t)
    \end{aligned}
\end{align}
where $N=\sqrt{N_AN_P},\;f_P=N_P/N_A,\;v=\sqrt{v_Av_P},\;v_A=k/N_A$, and $v_P=\kappa/N_P$. This will eventually result in degree dependent terms $a_v(t)$ and $p_v(t)$ given by
\begin{align}
    \begin{aligned}
         \int &\mathcal D[\dots]\mathcal M_A\mathcal M_P\exp \left(i \sum_i \int dt \; P_i(t) \psi_i(t) \right) \exp \left(i \sum_i \int dt \; A_i(t) \phi_i(t) \right)\\
         &\times\exp\Biggr\lbrace i\int dt\sum_n\sum_i\left[a_n\hat b_n(-v_{P_i})^n\hat a_i+a_n{\hat d_n}(-v_{A_i})^n\hat p_i+\int dt'\left(a_n{\hat k_n}(-v_{P_i})^n\hat a_i\hat a_i'+a_n\hat g_n(-v_{A_i})^n\hat p_i\hat p_i'\right)\right]\Biggr \rbrace\\
         &\times\exp\Biggr\lbrace i\int dt\int dt'\sum_n\sum_i\left[a_n{\hat y_n}(-v_{A_i})^n\hat p_iA_i'+a_n{\hat z_n}(-v_{P_i})^n\hat a_iP_i'\right]\Biggr \rbrace\times({\rm similar\; stuff\;in\;beta})
    \end{aligned}
\end{align}
So the only terms that we get will be like
\begin{align}
    \begin{aligned}
       a_v(t)&=\sum_n\left(\frac{a_n}{\langle v_A\rangle}m_PB_n(-v_P)^n+\eta_n+\frac{a_n}{\langle v\rangle\sqrt{f_P}}s_Ps_Ac_\gamma\int dt' (-v_P)^ny_n(t,t')P_v(t')\right)\\
        {\rm and}\;\; p_v(t)&=\sum_n\left(\frac{a_n}{\langle v_P\rangle}m_AD_n(-v_A)^n+\xi_n+\frac{a_n\sqrt{f_P}}{\langle v\rangle}s_As_Pc_\gamma\int dt' (-v_A)^nz_n(t,t')A_u(t')\right)\\
        {\rm with} \;\;\langle \eta_n\eta_n'\rangle&=a_n\frac{s_P^2}{\langle v_A\rangle}(-v_P)^nK_n \;\;\;\;\;\;{\rm and} \;\;\;\;\;\;\langle\xi_n\xi_n'\rangle=a_n\frac{s_A^2}{\langle v_P\rangle}(-v_A)^nG_n
    \end{aligned}
\end{align}
 Consider just the first summation of the first expression,
 \begin{align}
     \begin{aligned}
         \sum_n\frac{a_n^2}{\langle v_A\rangle}m_P\langle v_A^n A_v\rangle_v(-v_P)^n&=\Bigg\langle \sum_n\frac{a_n^2}{\langle v_A\rangle}m_P A_v(-v_Av_P)^n\Bigg\rangle_A\\
         &=\frac{m_P}{\langle v_A\rangle}\Bigg\langle P(v_Av_P)A_v\Bigg\rangle_A
     \end{aligned}
 \end{align}
 which lets us rewrite the infinite sum more easily as 
\begin{align}
    \begin{aligned}
        a_v(t)&=m_PB_v(t)+\eta_v(t)+\sigma_P\sigma_Ac_\gamma\int dt'y_v(t,t')P_v(t'),\\
        B_v(t)&=\frac1{\langle v_A\rangle}\int dv_A \rho(v_A)\Big\langle P(v_Av_P)A_v(t)\Big\rangle,\\
        \langle \eta_v\eta_v'\rangle&=\frac{\sigma_P^2}{\langle v_A\rangle}\int dv_A \rho(v_A)\Big\langle P(v_Av_P)A_v(t)A_v(t')\Big\rangle,\\
        y_v(t,t')&=\frac1{\langle v_a\rangle}\int dv_A\rho(v_A)\Big\langle P(v_Av_P)\frac{\delta A_v(t)}{\delta\xi_v(t')}\Big\rangle.
    \end{aligned}
\end{align}
and
\begin{align}
    \begin{aligned}
        p_v(t)&=m_AD_v(t)+\xi_v(t)+\sigma_P\sigma_Ac_\gamma\int dt'z_v(t,t')A_v(t'),\\
        D_v(t)&=\frac1{\langle v_P\rangle}\int dv_P \rho(v_P)\Big\langle P(v_Av_P)P_v(t)\Big\rangle,\\
        \langle \xi_v\xi_v'\rangle&=\frac{\sigma_A^2}{\langle v_P\rangle}\int dv_P \rho(v_P)\Big\langle P(v_Av_P)P_v(t)P_v(t')\Big\rangle,\\
        z_v(t,t')&=\frac1{\langle v_P\rangle}\int dv_P\rho(v_P)\Big\langle P(v_Av_P)\frac{\delta P_v(t)}{\delta\eta_v(t')}\Big\rangle.
    \end{aligned}
\end{align}
which are to be inserted in the final equations
\begin{align}
    \begin{aligned}
        \dot A_v(t)&=A_v(t)\left(1-A_v(t)-\mu_AM_A-\epsilon_A(t)-\int dt'Q_A(t,t')A_v(t')+\frac{p_v(t)}{1+h_Ap_v(t)}\right)\\
        \dot P_v(t)&=P_v(t)\left(1-P_v(t)-\mu_PM_P-\epsilon_P(t)-\int dt'Q_P(t,t')P_v(t')+\frac{a_v(t)}{1+h_Pa_v(t)}\right)
    \end{aligned}
\end{align}
Given that we have a bipartite network, we need to keep in mind that the degree distributions for animals and plants are different. As proposed in \cite{patil2026dynamical}, the inverse hidden degree would have to calculated by doing the following iteratively
\begin{align}
v_A&=\int d\tilde v_P\tilde \rho(\tilde v_P)P(\tilde v_A\tilde v_P)    \\
v_P&=\int d\tilde v_A\tilde \rho(\tilde v_A)P(\tilde v_A\tilde v_P)  \\
\tilde v_A&=\frac{v_A}{\int d\tilde v_P\tilde \rho(\tilde v_P)\frac{\tilde v_P}{\tilde v_A\tilde v_P+\sqrt{c_Ac_P}}}\\
\tilde v_P&=\frac{v_P}{\int d\tilde v_A\tilde \rho(\tilde v_A)\frac{\tilde v_A}{\tilde v_A\tilde v_P+\sqrt{c_Ac_P}}},
\end{align}
which will give the exact probability of connection of two nodes with given degrees.

\subsection{Effect of Nestedness on equilibrium}

Let's consider the simpler form of the system
\begin{align}
    \begin{aligned}
        \dot A_i&=A_i(1-A_i-\sum_j \alpha_{ij}A_j+\sum_jC_{ij} \beta_{ij}P_j)\\
        \dot P_i&=P_i(1-P_i+\sum_j\gamma_{ij}C_{ji}A_j-\sum_j\chi_{ij}P_j)
    \end{aligned}
\end{align}
Now let's say animals equilibriate quickly, giving
\begin{align}
    \begin{aligned}
        \vec A^*=(\mathbb1+\alpha)^{-1}(1+C\beta)\vec P\approx (\mathbb1-\alpha)(1+C\beta)\vec P
    \end{aligned}
\end{align}
where we use the fact that $\alpha$ has $1/N$ scalings. This gives
\begin{align}
    \dot P_i&=P_i(1-P_i-\sum_j\chi_{ij}P_j+\sum_j\gamma_{ij}C_{ji}\sum_k\sum_l(\delta_{jk}-\alpha_{jk})(\delta_{kl}+C_{kl}\beta_{kl})P_l)\\
    \dot P_i&=P_i(1-P_i-\sum_j\chi_{ij}P_j+\sum_{jkl}\gamma_{il}C_{li}(\delta_{lk}-\alpha_{lk})(\delta_{kj}+C_{kj}\beta_{kj})P_j)\\
    \implies \chi_{ij}'&=\chi_{ij}-\sum_{kl}\gamma_{il}C_{li}(\delta_{lk}-\alpha_{lk})(\delta_{kj}+C_{kj}\beta_{kj})\\
    &=\chi_{ij}-\gamma_{ij}C_{ji}-\sum_{k}\gamma_{ik}C_{ki}C_{kj}\beta_{kj}+\sum_l\gamma_{il}C_{li}\alpha_{lj}+\sum_{kl}\gamma_{il}C_{li}\alpha_{lk}C_{kj}\beta_{kj}
\end{align}
So it's as if we're in a regular GLV with a modified interaction matrix. The disorder average of each element in the matrix is
\begin{align}
    \begin{aligned}
        \langle \chi_{ij}'\rangle
        &\approx \frac{\mu_\chi}{N}-\frac{\mu_\gamma}{N}P_{ji}-\frac{\mu_\gamma^2}{N}n_{ij}+\frac{\mu_\alpha\mu_\gamma}{N^2}k_i+\frac{\mu_\alpha}{N}\frac{\mu_\gamma^2 k_ik_j}{N^2}
    \end{aligned}
\end{align}
where we drop terms which are too small but have to keep the terms that are in order $k/N$ cuz they can be finite. Here we define $n_{ij}$ as the overlap  of two nodes,
\begin{align}
    n_{ij}=\frac1N\sum_k C_{ki}C_{kj},
\end{align}
the average of $n_{ij}/{\rm min}(k_i,k_j)$ over all pairs is the definition of the nestedness of the network. Now
\begin{align}
    \begin{aligned}
        \langle \chi_{ij}'\chi_{ij}'\rangle&=\langle (\chi_{ij}-\gamma_{ij}C_{ji}-\sum_{k}\gamma_{ik}C_{ki}C_{kj}\beta_{kj}+\sum_k\gamma_{ik}C_{ki}\alpha_{kj}+\sum_{kl}\gamma_{il}C_{li}\alpha_{lk}C_{kj}\beta_{kj})\\&\qquad\times(\chi_{ij}-\gamma_{ij}C_{ji}-\sum_{m}\gamma_{im}C_{mi}C_{mj}\beta_{mj}+\sum_m\gamma_{im}C_{mi}\alpha_{mj}+\sum_{mn}\gamma_{in}C_{ni}\alpha_{nm}C_{mj}\beta_{mj})\rangle
    \end{aligned}
\end{align}
which we split into all its terms
\begin{align}
    \begin{aligned}
        \langle \chi_{ij}^2\rangle&=\frac{\mu_\chi^2}{N^2}+\frac{\sigma_\chi^2}{N}\\
        \langle \chi_{ij}\gamma_{ij}C_{ji}\rangle&=\frac{\mu_\gamma\mu_\chi}{N^2}P_{ji}\\
        \sum_m\langle \chi_{ij}\gamma_{im}C_{mi}C_{mj}\beta_{mj}\rangle&=\mu_\chi(\frac{\mu_\gamma^2}{N^2}+\frac{\sigma_\gamma^2c_\gamma}{N}\delta_{ij})n_{ij}\\
        \sum_m\langle\chi_{ij}\gamma_{im}C_{mi}\alpha_{mj}\rangle&=\frac{\mu_\gamma\mu_\chi^2}{N^2}\frac{k_i}{N}\\
        \sum_{mn}\langle \chi_{ij}\gamma_{in}C_{ni}C_{mj}\beta_{mj}\alpha_{nm}\rangle&=\frac{\mu_\chi\mu_\alpha}{N^2}(\mu_\gamma^2\frac{k_ik_j}{N^2}+\delta_{nm}\delta_{ij}\sigma_\gamma^2c_\gamma P_{ij})\\
        \langle\gamma_{ij}^2C_{ji}^2\rangle&=(\frac{\mu_\gamma^2}{N^2}+\frac{\sigma_\gamma^2}{N})P_{ji}\\
        \sum_m\langle \gamma_{ij}C_{ji}\gamma_{im}C_{mi}C_{mj}\beta_{mj}\rangle&=\frac{\mu_\gamma^3}{N^3}+\frac{\sigma_\gamma^2\mu_\gamma}{N^2}P_{ji}P_{jj}+\delta_{ij}\frac{k_i}{N}\frac{\sigma_\gamma^2\mu_\gamma c_\gamma}{N}\\        
        \sum_m\langle \gamma_{ij}C_{ji}\gamma_{im}C_{mi}\alpha_{mj}\rangle&=\frac{\mu_\alpha\mu_\gamma^2}{N^3}+\frac{\mu_\alpha\sigma_\gamma^2}{N^2}P_{ji}\\        
        \sum_{mn}\langle \gamma_{ij}C_{ji}\gamma_{in}C_{ni}\alpha_{nm}C_{mj}\beta_{mj}\rangle&=\mathcal O(1/N^2)\\        \sum_{km}\langle\gamma_{ik}C_{ki}C_{kj}\beta_{kj}\gamma_{im}C_{mi}C_{mj}\beta_{mj}\rangle&\approx\frac{\sigma_\gamma^4n_{ij}}{N}+\frac{\sigma_\gamma^4c_\gamma^2k_i^2}{N^2}\delta_{ij}\\
        \sum_{km}\langle \gamma_{ik}C_{ki}C_{kj}\beta_{kj}\gamma_{im}C_{mi}\alpha_{mj}\rangle&\approx\mathcal O(1/N^2)\\        \sum_{kmn}\langle\gamma_{ik}C_{ki}C_{kj}\beta_{kj}\gamma_{in}C_{ni}C_{mj}\beta_{mj}\alpha_{nm}\rangle&\approx\mathcal O(1/N^2)\\
        \sum_{km}\langle \gamma_{ik}C_{ki}\alpha_{kj}\gamma_{im}C_{mi}\alpha_{mj}\rangle&=\frac{\sigma_\alpha^2\sigma_\gamma^2}{N}\frac{k_i}{N}\\
        \sum_{kmn}\langle \gamma_{ik}C_{ki}\alpha_{kj}\gamma_{in}C_{ni}\alpha_{nm}C_{mj}\beta_{mj}\rangle&\approx\mathcal O(1/N^2)\\
        \sum_{klmn}\langle \gamma_{il}C_{li}\alpha_{lk}C_{kj}\beta_{kj}\gamma_{in}C_{ni}\alpha_{nm}C_{mj}\beta_{mj}\rangle        &\approx \frac{\sigma_\alpha^2\sigma_\gamma^4 k_ik_j}{N^3}
    \end{aligned}
\end{align}
giving that the effective competition matrix has a covariance
\begin{align}
    \begin{aligned}
        \langle \chi_{ij}'\chi_{ij}'\rangle_c&=\frac{\sigma_\chi^2}{N}+\frac{\sigma_\gamma^2}{N}P_{ji}+\frac{\sigma_\gamma^4n_{ij}}{N}+\delta_{ij}\sigma_\gamma^4c_\gamma^2\frac{k_i^2}{N^2}+\frac{\sigma_\alpha^2\sigma_\gamma^2}{N}\frac{k_i}{N}+\frac{\sigma_\gamma^2\sigma_\gamma^4}{N}\frac{k_ik_j}{N^2}.
    \end{aligned}
\end{align}
Similarly we can compute the covariance between a term and its transpose element,
\begin{align}
    \begin{aligned}
        \langle \chi_{ij}\chi_{ji}\rangle&=\frac{\mu_\chi^2}{N^2}+\frac{\sigma_\chi^2}{N}c_\chi\\
        \langle \chi_{ij}\gamma_{ji}C_{ii}\rangle&=\frac{\mu_\gamma\mu_\chi}{N^2}P_{ji}\\
        \sum_m\langle \chi_{ji}\gamma_{im}C_{mi}C_{mj}\beta_{mj}\rangle&=\mu_\chi(\frac{\mu_\gamma^2}{N^2}+\frac{\sigma_\gamma^2c_\gamma}{N}\delta_{ij})n_{ij}\\
        \sum_m\langle\chi_{ji}\gamma_{im}C_{mi}\alpha_{mj}\rangle&=\frac{\mu_\gamma\mu_\chi^2}{N^2}\frac{k_i}{N}\\
        \sum_{mn}\langle \chi_{ji}\gamma_{in}C_{ni}C_{mj}\beta_{mj}\alpha_{nm}\rangle&=\frac{\mu_\chi\mu_\alpha}{N^2}(\mu_\gamma^2\frac{k_ik_j}{N^2}+\delta_{nm}\delta_{ij}\sigma_\gamma^2c_\gamma P_{ij})\\
        \langle\gamma_{ij}\gamma_{ji}C_{ji}C_{ij}\rangle&=\frac{\mu_\gamma^2}{N^2}P_{ji}P_{ij}\\
        \sum_m\langle \gamma_{ji}C_{ij}\gamma_{im}C_{mi}C_{mj}\beta_{mj}\rangle&=\mathcal O(1/N^2)\\        
        \sum_m\langle \gamma_{ji}C_{ij}\gamma_{im}C_{mi}\alpha_{mj}\rangle&=\mathcal O(1/N^2)\\        
        \sum_{mn}\langle \gamma_{ij}C_{ji}\gamma_{in}C_{ni}\alpha_{nm}C_{mj}\beta_{mj}\rangle&=\mathcal O(1/N^2)\\        \sum_{km}\langle\gamma_{ik}C_{ki}C_{kj}\beta_{kj}\gamma_{jm}C_{mj}C_{mi}\beta_{mi}\rangle&\approx\frac{\sigma_\gamma^4c_\gamma^2n_{ij}}{N}+\frac{\sigma_\gamma^4c_\gamma^2k_i^2}{N^2}\delta_{ij}\\
        \sum_{km}\langle \gamma_{ik}C_{ki}C_{kj}\beta_{kj}\gamma_{jm}C_{mj}\alpha_{mi}\rangle&\approx\mathcal O(1/N^2)\\        \sum_{kmn}\langle\gamma_{jk}C_{kj}C_{ki}\beta_{ki}\gamma_{in}C_{ni}C_{mj}\beta_{mj}\alpha_{nm}\rangle&\approx\mathcal O(1/N^2)\\
        \sum_{km}\langle \gamma_{jk}C_{kj}\alpha_{ki}\gamma_{im}C_{mi}\alpha_{mj}\rangle&=\frac{\sigma_\alpha^2\sigma_\gamma^2}{N}\frac{k_i}{N}\delta_{ij}\\
        \sum_{kmn}\langle \gamma_{ik}C_{ki}\alpha_{kj}\gamma_{in}C_{ni}\alpha_{nm}C_{mj}\beta_{mj}\rangle&\approx\mathcal O(1/N^2)\\
        \sum_{klmn}\langle \gamma_{il}C_{li}\alpha_{lk}C_{kj}\beta_{kj}\gamma_{jn}C_{nj}\alpha_{nm}C_{mi}\beta_{mi}\rangle        &\approx\mathcal O(1/N^2)
    \end{aligned}
\end{align}
which can be condensed into the leading order terms \begin{align}
    \begin{aligned}
        \langle \chi_{ij}'\chi_{ji}'\rangle&=\frac{\sigma_\chi^2c_\chi}{N}+\frac{\sigma_\gamma^4c_\gamma ^2n_{ij}}{N}+\delta_{ij}\sigma_\gamma^4c_\gamma^2\frac{k_i^2}{N^2}.
    \end{aligned}
\end{align}
Thus the effective interaction matrix has the statistics
\begin{align}
    \begin{aligned}
        \langle \chi_{ij}'\rangle&= \frac{\mu_\chi}{N}-\frac{\mu_\gamma}{N}P_{ji}-\frac{\mu_\gamma^2}{N}f_{ij}{\rm min}(v_i,v_j)+\frac{\mu_\alpha\mu_\gamma}{N}v_i+\frac{\mu_\alpha}{N}\mu_\gamma^2 v_iv_j,\\
        \langle \chi_{ij}'\chi_{ij}'\rangle&=\frac{\sigma_\chi^2}{N}+\frac{\sigma_\gamma^2}{N}P_{ji}+\frac{\sigma_\gamma^4}{N}f_{ij}{\rm min}(v_i,v_j)+\delta_{ij}\sigma_\gamma^4c_\gamma^2v_i^2+\frac{\sigma_\alpha^2\sigma_\gamma^2}{N}v_i+\frac{\sigma_\gamma^2\sigma_\gamma^4}{N}v_iv_j,\\
        \langle\chi_{ij}'\chi_{ji}'\rangle&=\frac{\sigma_\chi^2c_\chi}{N}+\frac{\sigma_\gamma^4c_\gamma ^2}{N}f_{ij}{\rm min}(v_i,v_j)+\delta_{ij}\sigma_\gamma^4c_\gamma^2v_i^2
    \end{aligned}
\end{align}
where we replace the overlap $n$ with the fraction overlap $f_{ij}\times {\rm min} (v_i,v_j)$, where ${\rm max}(v_i,v_j)<f_{ij}<1$. Which is to say the expected neutral overlap is the product of the connectivities while the maximum nested overlap is equal to the connectivity of the node with lower connectivity. The average $F=\frac{1}{N^2}\sum_{ij}f_{ij}$ gives the total nestedness of the system, while the average of the variances of the terms of the effective competition matrix gives how large we expect its spectral radius to be. This term depends on the nestedness in the way
\begin{align}
    X\propto\sum_{ij}f_{ij}{\rm min}(v_i,v_j).
\end{align}
We need to find the relation between $X$ and $F$ to establish at least a qualitative understanding of the effect of nestedness on the stability of an ecosystem.
To this end, imagine stretching out the "matrices" $f_{ij}$ and ${\rm min}(v_i,v_j)$ into "vectors". The problem then is to find a "vector" $\vec f$ that has maximum overlap with $\vec 1$ -- increasing nestedness -- while minimizing overlap with $\vec m (= {\rm min} (u,v))$, i.e. reducing the total covariance of the terms of the stability matrix. Now highly heterogeneous networks have a certain amount of nestedness by default, let's call that $\vec f_0$. If we change each overlap by the vector $\vec \epsilon$ without letting the degrees change, we have
\begin{align}
    \delta F=\vec \epsilon\cdot 1,\qquad \qquad
    \delta X=\vec \epsilon\cdot \vec m.
\end{align}
Without any constraints on $\vec\epsilon$, it is theoretically possible to have a positive $\delta F$ and a negative $\delta X$, by choosing to have large increases in nestedness where $\vec m$ is small and reductions where it is large. However, if we naïvely choose an $\vec \epsilon$ with all positive values -- increasing all overlaps -- it will unequivocally lead to more disordered effective interaction matrix, and thus a less stable system.

\subsection{Nestedness in dynamics}

Extending the above analysis to the dynamical system using a HDMFT framework, consider once again the generating function of the model,
\begin{align}
\begin{aligned}
Z[\boldsymbol{\psi},\boldsymbol{\phi}]=\int & \mathcal{D}[\dots] \int \prod_i  \frac{d a_i d\hat{a}_i}{2 \pi}\frac{d p_i d\hat{p}_i}{2 \pi}  \exp \Biggl \lbrace i \sum_i \int dt\hat{a}_i \left(a_i -\sum_k C_{ik}\gamma^P_{ik}A_k \right) \Biggr \rbrace 
\exp \Biggl \lbrace i \sum_i \int dt\hat{p}_i \left(p_i -\sum_kC_{ki}\gamma_{ik}^A P_k \right) \Biggr \rbrace \\\times
& \exp \Biggl \lbrace i \sum_i \int dt  \hat{P}_i(t) \left[ \frac{\dot{P}_i}{P_i}-\left(\alpha -\chi P_i-\sum_{j} \beta_{ij}^{(P)} P_j+\frac{a_i}{1+h^{(P)} a_i}  +\epsilon_i \right) \right] \Biggr \rbrace  \\\times
& \exp \Biggl \lbrace i \sum_i \int dt  \hat{A}_i(t) \left[ \frac{\dot{A}_i}{A_i}-\left(\alpha -\chi A_i-\sum_{j } \beta_{ij}^{(A)} A_j+\frac{p_i}{1+h^{(A)} p_i}  +\xi_i \right) \right] \Biggr \rbrace 
 \\& \times\exp \left(i \sum_i \int dt \; P_i(t) \psi_i(t) \right) \exp \left(i \sum_i \int dt \; A_i(t) \phi_i(t) \right) \prod_i\\
 =\int &\mathcal D[\dots]\mathcal M_A\mathcal M_P\exp \left(i \sum_i \int dt \; P_i(t) \psi_i(t) \right) \exp \left(i \sum_i \int dt \; A_i(t) \phi_i(t) \right)\\
 &\times \exp\Biggr\lbrace- i\sum_i\int dt\left[\hat a_i\sum_j C_{ij}\gamma_{ij}^PA_j+\hat p_i\sum_jC_{ji}\gamma_{ij}^AP_j\right]\Biggr\rbrace
\end{aligned}
\end{align}
where
\begin{align}
    \begin{aligned}
        \mathcal M_A&=\exp\Biggl \lbrace i \sum_i  \int dt  \left[ \hat{A}_i(t)\left( \frac{\dot{A}_i}{A_i}-1 + A_i+\beta^{\rm (eff) A}_i-\frac{p_i}{1+h^{(A)} p_i} -\xi_i \right)  +\hat a_i(t)a_i(t)\right]\Biggr \rbrace \\
        \mathcal M_P&=\exp\Biggl \lbrace i \sum_i  \int dt  \left[ \hat{P}_i(t)\left( \frac{\dot{P}_i}{P_i}-1 + P_i++\beta^{\rm (eff) P}_i-\frac{a_i}{1+h^{(P)} a_i} -\epsilon_i \right)  +\hat p_i(t)p_i(t)\right]\Biggr \rbrace 
    \end{aligned}
\end{align}
Now consider only the average over $C$ and $\gamma$, this time assuming that different edges might be correlated
\begin{align}
    \begin{aligned}
        &\Bigg\langle\overline{\exp\Biggr\lbrace-i\sum_{ij}C_{ij}\left(\hat a_i\gamma_{ij}^PA_j+\hat p _j\gamma_{ji}^AP_i\right)\Bigg\rbrace}\Bigg\rangle=\Bigg\langle\overline{\exp(\sum_{ij}C_{ij}Z_{ij})}\Bigg\rangle\\
        &=\Bigg\langle\overline{1+\sum_{ij}C_{ij}Z_{ij}+\frac{1}{2}\sum_{ijkl}C_{ij}C_{kl}Z_{ij}Z_{kl}'+\dots}\Bigg\rangle\\
        &\approx\Bigg\langle1+\sum_{ij}P(i,j)Z_{ij}+\frac{1}{2}\sum_{ijkl}P(ij,kl)Z_{ij}Z_{kl}'+\dots\Bigg\rangle\\
        &\approx \Bigg\langle\exp\left(\sum_{ij}P(i,j)Z_{ij}+\frac12\sum_{ijkl}[P(ij,kl)-P(i,j)P(k,l)]Z_{ij}Z_{kl}'+\dots\right)\Bigg\rangle\\
        &\approx \exp\left(-i\sum_{ij}P(i,j)\frac{\mu}{C}(\hat a_iA_j+\hat p_jP_i)-\frac12\sum_{ijl}f(ij,il)\frac{\mu^2}{C^2}(\hat a_iA_j\hat a_i'A_l'+\hat p_jP_i\hat a_i'A_l'+\hat a_iA_j\hat p_l'P_i'+\hat p_jP_i\hat p_l'P_i')\right.\\&\qquad\qquad\left.-\frac12\sum_{ijk}f(ij,kj)\frac{\mu^2}{C^2}(\hat a_iA_j\hat a_k'A_j'+\hat p_jP_i\hat a_k'A_j'+\hat a_iA_j\hat p_j'P_k'+\hat p_jP_i\hat p_j'P_k')\right.\\&\qquad\qquad\left.-\frac12\sum_{ij}P(i,j)\frac{\sigma^2}{C}(\hat a_iA_j\hat a_i'A_j'+c_\gamma\hat a_iA_j\hat p_j'P_i'+c_\gamma\hat p_jP_i\hat a_i'A_j'+\hat p_jP_i\hat p_j'P_i')\right)
        \end{aligned}
\end{align}
where $f(ij,kl)=P(ij,kl)-P(i,j)P(k,l)$. This object is assumed to be only non-zero for the nestedness like correlations, leaving only this term present and all higher order terms zero as well. This is to say that we choose networks that only have the correlations naturally present due to the degree distribution, plus additional ``nestedness-like'' overlaps.

When $f$ is positive it means there is a positive correlation between two edges, as in they're more likely to be present simultaneously than if they were not related. A negative $f$ implies $k-l$ being present reduces the chances that $i-j$ will exist. More positive $f$'s will imply higher nestedness. Now let's define the order paramaters
\begin{align}
    \begin{aligned}
        b_i&=\frac1C\sum_jP(k_i,k_j)A_j&\qquad d_i&=\frac1C\sum_jP(k_j,k_i)P_j\\
        B_i&=\frac1C\sum_jP(k_i,k_j)A_jA_j'&\qquad D_i&=\frac1C\sum_jP(k_j,k_i)P_jP_j'\\
        K_i&=\frac1{C^2}\sum_{jl}f(k_ik_j,k_ik_l)A_jA_l'&\qquad G_i&=\frac1{C^2}\sum_{jl}f(k_jk_i,k_lk_i)P_jP_l'\\
        z_{i}&=\frac1{C^2}\sum_{jl}f(k_ik_j,k_ik_l)\hat p_jA_l'&\qquad y_{i}&=\frac1{C^2}\sum_{jl}f(k_jk_i,k_lk_i)\hat a_l P_j'\\
        Z_{jl}&=\frac1{C}\sum_{i}f(k_ik_j,k_ik_l)\hat a_i'P_i&\qquad Y_{jl}&=\frac1{C}\sum_{i}f(k_jk_i,k_lk_i)\hat p_i' A_i\\
        U_{jl}&=\frac1{C}\sum_{i}f(k_ik_j,k_ik_l)P_iP_i'&\qquad V_{jl}&=\frac1{C}\sum_{i}f(k_jk_i,k_lk_i)A_iA_i'\\
        X_{1_i}&=\frac1{C}\sum_{j}P(k_i,k_j)\hat a_jP_j'&\qquad X_{2_i}&=\frac1{C}\sum_{j}P(k_i,k_j)A_j\hat p_j'
        \\
    \end{aligned}
\end{align}

which can be inserted into the generating function to separate the indices, giving
 \begin{align}
     \begin{aligned}
         \int &\mathcal D[\dots]\mathcal M_A\mathcal M_P\exp \left(i \sum_i \int dt \; P_i(t) \psi_i(t) \right) \exp \left(i \sum_i \int dt \; A_i(t) \phi_i(t) \right)\\
         \times&\exp\left(-i\sum_{i}\mu(\hat a_i b_i+\hat p_id_i)-\frac{\mu^2}2\sum_{i}[\hat a_i\hat a_i'K_i(t,t')+P_i'\hat a_iz_{i}(t',t)]-\frac{\mu^2}{2C}\sum_{jl}[\hat p_j\hat p_l'U_{jl}(t,t')+A_l'\hat p_jZ_{jl}(t,t')]\right.\\&\qquad\qquad\left.-\frac{\mu^2}2\sum_{i}[\hat p_iA_i'y_{i}(t',t)+\hat p_i\hat p_i'G_i(t,t')]-\frac{\mu^2}{2C}\sum_{jl}[V_{jl}(t,t')\hat a_j\hat a_l'+\hat a_jP_lY_{jl}(t,t')]\right.\\&\qquad\qquad\left.-\frac{\sigma^2}2\sum_{i}(\hat a_i\hat a_i'B_i(t,t')+c_\gamma\hat a_iP_i'X_{1_i}(t',t)+c_\gamma A_i\hat p_i'X_{2_i}(t,t') +\hat p_i\hat p_i'D_i(t,t'))\right)\\
        = \int &\mathcal D[\dots]\mathcal M_A\mathcal M_P\exp \left(i \sum_i \int dt \; P_i(t) \psi_i(t) \right) \exp \left(i \sum_i \int dt \; A_i(t) \phi_i(t) \right)\\
         \times&\exp\left(-i\sum_{i}\int dt\hat a_i(t)\mu b_i(t)-i\sum_i\int dt\hat p_i(t) \mu d_i(t)\right.\\&\qquad-\frac12\sum_{ij}\int dtdt'\hat a_i(t)\hat a_j(t')[\sigma^2B_i(t,t')\delta_{ij}+\mu^2K_i(t,t')\delta_{ij}+\frac{\mu^2}{C}U_{ij}(t,t')]\\
         &\qquad-\frac12\sum_{ij}\int dtdt'\hat p_i(t)\hat p_j(t')[\sigma^2D_i(t,t')\delta_{ij}+\mu^2G_i(t,t')\delta_{ij}+\frac{\mu^2}{C}V_{ij}(t,t')]\\
         &\qquad-\frac12\sum_{ij}\int dtdt'\hat a_i(t)P_j(t')[\sigma^2c_\gamma X_{2_i}(t,t')\delta_{ij}+\mu^2z_{i}(t',t)\delta_{ij}+\frac{\mu^2}{C}Y_{ij}(t',t)]\\
         &\left.\qquad-\frac12\sum_{ij}\int dtdt'\hat p_i(t)A_i(t')[\sigma^2c_\gamma X_{1_i}(t',t)\delta_{ij}+\mu^2y_{i}(t',t)\delta_{ij}+\frac{\mu^2}{C}Z_{ij}(t,t')]\right)\\
         =\int &\mathcal D[\dots]\mathcal M_A\mathcal M_P\exp\left(i\sum_i\int dtP_i(t)\psi_i(t)\right)\exp\left(i\sum_i\int dtA_i(t)\phi_i(t)\right)\\
         \times &\exp\left(-i\sum_i\hat a_i\left[\mu b_i(t)+\eta^{(1)}_i(t)+\eta^{(2)}_i(t)+\int dt' R^{(1)}_{i}(t,t')P_i(t')+\int dt' R^{(2)}_{ij}(t,t')P_j(t')\right]\right)\\
         \times &\exp\left(-i\sum_i\hat p_i\left[\mu d_i(t)+\epsilon^{(1)}_i(t)+\epsilon^{(2)}_i(t)+\int dt' Q^{(1)}_{i}(t,t')A_i(t')+\int dt' Q^{(2)}_{ij}(t,t')A_j(t')\right]\right).
     \end{aligned}
 \end{align}
 Here the various correlations and responses are defined as
 \begin{align}
     \begin{aligned}
     \langle v\rangle b_i(t)&=\langle P(ij)A_j(t)\rangle_j\\
    \langle v\rangle  d_i(t)&=\langle P(ji)P_j(t)\rangle_j\\
        \langle v\rangle  \langle \eta^{(1)}_i(t)\eta^{(1)}_i(t')\rangle&=\langle v\rangle \sigma^2B_i(t,t')=\sigma^2\langle P(i,j)A_j(t)A_j(t')\rangle_j\\
        \langle v\rangle  \langle \epsilon^{(1)}_i(t)\epsilon^{(1)}_i(t')\rangle&=\langle v\rangle \sigma^2D_i(t,t')=\sigma^2\langle P(j,i)P_j(t)P_j(t')\rangle_j\\
        \langle v\rangle  \langle \eta^{(2)}_i(t)\eta^{(2)}_j(t')\rangle&=\langle v\rangle \frac{\mu^2}{C}U_{ij}(t,t')+\langle v\rangle \mu^2K_i(t,t')\delta_{ij}=\frac{\mu^2}{C}\langle f(lj,li)P_l(t)P_l(t')\rangle_{l}+\mu^2\langle f(ik,il)A_k(t)A_l(t')\rangle_{kl}\delta_{ij}\\
        \langle v\rangle  \langle \epsilon^{(2)}_i(t)\epsilon^{(2)}_j(t')\rangle&=\langle v\rangle \frac{\mu^2}{C}V_{ij}(t,t')+\langle v\rangle \mu^2G_i(t,t')\delta_{ij}=\frac{\mu^2}{C}\langle f(jl,il)A_l(t)A_l(t')\rangle_{l}+\mu^2\langle f(ki,li)P_k(t)P_l(t')\rangle_{kl}\delta_{ij}\\
        \langle v\rangle  R_i^{(1)}(t,t')&=\sigma^2c_\gamma \langle v\rangle X_{2_i}(t,t')=\sigma^2c_\gamma \Bigg \langle P(ji)\frac{\delta P_j(t)}{\delta \eta_j(t')}\Bigg\rangle_j\\
        \langle v\rangle  Q_i^{(1)}(t,t')&=\sigma^2c_\gamma \langle v\rangle X_{1_i}(t',t)=\sigma^2c_\gamma \Bigg\langle P(ij)\frac{\delta A_j(t)}{\delta \epsilon_j(t')}\Bigg\rangle_j\\
        \langle v\rangle  R_{ij}^{(2)}(t,t')&=\frac{\mu^2}{C}\langle v\rangle Y_{ij}(t,t')+\mu^2\langle v\rangle z_i(t',t)\delta_{ij}=\frac{\mu^2}{C}\Bigg\langle f(il,jl)\frac{\delta A_l(t)}{\delta \epsilon_l(t')}\Bigg\rangle_{l}+\mu^2\Bigg\langle f(ik,il)\frac{\delta A_l(t)}{\delta \epsilon_k(t')}\Bigg\rangle_{kl}\delta_{ij}\\
        \langle v\rangle  Q_{ij}^{(2)}(t,t')&=\frac{\mu^2}{C}\langle v\rangle Z_{ij}(t,t')+\mu^2\langle v\rangle y_i(t',t)\delta_{ij}=\frac{\mu^2}{C}\Bigg\langle f(li,lj)\frac{\delta P_l(t)}{\delta \eta_l(t')}\Bigg\rangle_{l}+\mu^2\Bigg\langle f(ki,li)\frac{\delta P_l(t)}{\delta \eta_k(t')}\Bigg\rangle_{kl}\delta_{ij}.
     \end{aligned}
 \end{align}
If we set the mean competitive interactions to zero, no non-linearities, and then take the ``mutualistic" nested interactions only (with arbitrary sign of the mean),
\begin{align}
    \begin{aligned}
        \dot A_i(t)&=A_i(t)(1-A_i(t)+\beta_i^{A}(t)+\mu d_i(t)+\epsilon^{(1)}_i(t)+\epsilon^{(2)}_i(t)+\int dt'Q_i^{(1)}(t,t')A_i(t')+\sum_j\int dt'Q_{ij}^{(2)}(t,t')A_j(t'))\\
        \dot P_i(t)&=P_i(t)(1-P_i(t)+\beta_i^{P}(t)+\mu b_i(t)+\eta^{(1)}_i(t)+\eta^{(2)}_i(t)+\int dt'R_i^{(1)}(t,t')P_i(t')+\sum_j\int dt'R_{ij}^{(2)}(t,t')P_j(t'))
    \end{aligned}
\end{align}
Taking HMFT assumptions and writing $f(ij,kl)$ as a function of the degrees, this is an effective equation in terms of the degrees of the nodes it is in, i.e.
\begin{align}
    \begin{aligned}
        \dot A_v(t)&=A_v(t)(1-A_v(t)+\beta_v^{A}(t)+\mu d_v(t)+\epsilon^{(1)}_v(t)+\epsilon^{(2)}_v(t)+\int dt'Q_v^{(1)}(t,t')A_v(t')+\int \rho(u)du\int dt'Q_{vu}^{(2)}(t,t')A_u(t'))\\
        \dot P_v(t)&=P_v(t)(1-P_v(t)+\beta_v^{P}(t)+\mu b_v(t)+\eta^{(1)}_v(t)+\eta^{(2)}_v(t)+\int dt'R_v^{(1)}(t,t')P_v(t')+\int \rho(u)du \int dt'R_{vu}^{(2)}(t,t')P_u(t'))
    \end{aligned}
\end{align}
with all parameter definitions rewritten in terms of degrees as well. Now performing a perturbative analysis of the previous equation and taking a fourier transform, we get
\begin{align}
    \begin{aligned}
        i\omega\delta  \tilde A_i&=A_i^*(-\delta \tilde A_i+\delta\tilde \beta_i^A+\delta \tilde \epsilon^{(1)}_i+\delta \tilde \epsilon^{(2)}_i+\tilde Q_i^{(1)}\delta \tilde A_i+\sum_j\tilde Q_{ij}^{(2)}\delta \tilde A_j)\\
        i\omega \delta \tilde \dot P_i&=P_i^*(-\delta \tilde P_i+\delta\tilde \beta_i^P+\delta \tilde \eta^{(1)}_i+\delta \tilde \eta^{(2)}_i+\tilde R_i^{(1)}\delta\tilde  P_i+\sum_j\tilde R_{ij}^{(2)}\delta\tilde  P_j)\\
        & \implies\\
        \sum_j\left(\left[\frac{i\omega}{A_i^*}+1-\tilde Q_i^{(1)}\right]\delta_{ij}-\tilde Q_{ij}^{(2)}  \right)\delta\tilde A_j&=\delta \tilde \epsilon^{(1)}_i+\delta \tilde \epsilon^{(2)}_i+\delta\tilde \beta_i^A+\delta\tilde h_1\\
        \sum_j\left(\left[\frac{i\omega}{P_i^*}+1-\tilde R_i^{(1)}\right]\delta_{ij}-\tilde R_{ij}^{(2)}  \right)\delta\tilde P_j&=\delta \tilde \eta^{(1)}_i+\delta \tilde \eta^{(2)}_i+\delta\tilde \beta_i^P+\delta\tilde h_2
    \end{aligned}
\end{align}

Let the matrices on the left be called $L$ and $M$ respectively. This gives
\begin{align}
    \begin{aligned}
        \frac{\langle \delta \tilde A_i^*\delta \tilde A_j\rangle}{\phi^A_i\phi^A_j}&=\sum_{kl}L_{ik}^{*-1}L_{jl}^{-1} \left(\delta_{kl}[\sigma_\beta^2\langle|\delta\tilde A^2_m|\rangle_m+\frac{\sigma_\gamma^2}{c}\langle P(km)|\delta\tilde P_m^2|\rangle_m+\frac{\mu_\gamma^2}{c}\langle f_{mk,nk}\delta \tilde P_m^*\delta \tilde P_n\rangle_{mn}]+\frac{\mu_\gamma^2}{c^2N}\langle f_{km,lm}|\delta \tilde A_m^2|\rangle_m +1\right)\\
        \frac{\langle \delta \tilde P_i^*\delta \tilde P_j\rangle}{\phi^P_i\phi^P_j}&=\sum_{kl}M_{ik}^{*-1}M_{jl}^{-1} \left(\delta_{kl}[\sigma_\beta^2\langle|\delta\tilde P^2_m|\rangle_m+\frac{\sigma_\gamma^2}{c}\langle P(mk)|\delta\tilde A_m^2|\rangle_m+\frac{\mu_\gamma^2}{c}\langle f_{km,kn}\delta \tilde A_m^*\delta \tilde A_n\rangle_{mn}]+\frac{\mu_\gamma^2}{c^2N}\langle f_{mk,ml}|\delta \tilde P_m^2|\rangle_m +1\right)
    \end{aligned}
\end{align}
Now these are both also higher order matrix equations,
\begin{align}
    \begin{aligned}
         \langle \delta \tilde A_i^*\delta \tilde A_j\rangle&=\phi^A_i\phi^A_j\langle\sum_{mn}\mathcal L^{(2)}_{ijmn} \delta \tilde A_m^*\delta \tilde A_n\rangle+\phi^A_i\phi^A_j\langle\sum_{mn}\mathcal L^{(1)}_{ijmn} \delta \tilde P_m^*\delta \tilde P_n\rangle+\mathcal L^0_{ij}\\
        \langle \delta \tilde P_i^*\delta \tilde P_j\rangle&=\phi^P_i\phi^P_j\langle\sum_{mn}\mathcal M^{(2)}_{ijmn} \delta \tilde P_m^*\delta \tilde P_n\rangle+\phi^P_i\phi^P_j\langle\sum_{mn}\mathcal M^{(1)}_{ijmn} \delta \tilde A_m^*\delta \tilde A_n\rangle+\mathcal M^0_{ij}
    \end{aligned}
\end{align}

where
\begin{align}
    \begin{aligned}
        c\mathcal L^{(1)}_{ijmn}&=\sum_{k}L^{*-1}_{ik}L^{-1}_{jk}(\sigma_\gamma^2P(km)\delta_{mn}+\mu_\gamma^2f(mk,nk))\\
        \mathcal L^{(2)}_{ijmn}&=\sum_{kl}L^{*-1}_{ik}L^{-1}_{jl}(\sigma_\beta^2+\frac{\mu_\gamma^2}{c^2N}f(km,lm))\delta_{mn}\\
        c\mathcal M^{(1)}_{ijmn}&=\sum_{k}M^{*-1}_{ik}M^{-1}_{jk}(\sigma_\gamma^2P(mk)\delta_{mn}+\mu_\gamma^2f(km,kn))\\
        \mathcal M^{(2)}_{ijmn}&=\sum_{kl}M^{*-1}_{ik}M^{-1}_{jl}(\sigma_\beta^2+\frac{\mu_\gamma^2}{c^2N}f(mk,ml)\delta_{mn})\\
        \mathcal L^0_{ij}&=\sum_{kl}L^{*-1}_{ik}L^{-1}_{jl}\\
        \mathcal M^0_{ij}&=\sum_{kl}M^{*-1}_{ik}M^{-1}_{jl}
    \end{aligned}
\end{align}
which can be finally reduced to a single expression
\begin{align}
    \sum_{mn}\langle \delta_{im}\delta_{jn}-\phi^P_i\phi^P_j\mathcal M_{ijmn}^{(2)}\rangle\delta \tilde P_m^*\delta \tilde P_n&=\phi^P_i\phi^P_j\sum_{mnxy}\langle \mathcal M^{(1)}_{ijmn}\left( \delta_{mx}\delta_{ny}-\phi^A_m\phi^A_n\mathcal L_{mnxy}^{(2)}\right)^{-1}(\phi_m^A\phi_n^A\mathcal L^{(1)}_{mnxy}\delta \tilde P_x^*\delta \tilde P_y+\mathcal L^0 )\rangle +\mathcal M^0.
\end{align}
The matrix acting on the $\delta P\delta P$ vector here is 
\begin{align}
    ((\mathbb1\times\mathbb 1)-(\phi^P\times \phi^P)\mathcal M^{(2)})-(\phi^P\times \phi^P)\mathcal M^{(1)}[(\mathbb1\times\mathbb1)-(\phi^A\times\phi^A)\mathcal L^{(2)}]^{-1}((\phi^A\times\phi^A) \mathcal L^{(1)})
\end{align}
which needs to be non-singular for the perturbations to be finite. This can be written as
\begin{align}
    \lambda\left[\begin{pmatrix}
        \phi\times\phi \mathcal L^{(2)}&\phi\times\phi \mathcal L^{(1)}\\\phi\times\phi \mathcal M^{(1)}&\phi\times\phi \mathcal M^{(2)}
    \end{pmatrix}\right]<1,
\end{align}
where by $\lambda$ we mean the spectral radius of the matrix.

Now assuming that the mean mutualistic interaction, nestedness, and interactions correlation strengths are small enough, we could write in the large time limit ($\omega\rightarrow0$),
\begin{align}
    \begin{aligned}
        L^{-1}_{ij}&\approx(1+\tilde Q^{(1)}_i)\delta_{ij}+\tilde Q^{(2)}_{ij}\\
        M^{-1}_{ij}&\approx(1+\tilde R^{(1)}_i)\delta_{ij}+\tilde R^{(2)}_{ij}
    \end{aligned}
\end{align}

giving
\begin{align}
    \begin{aligned}
        c\mathcal L^{(1)}_{ijmn}&\approx\mu_\gamma^2f(mi,ni)\delta_{ij}+[2\tilde Q_i^{(1)}+2\tilde Q_{ij}^{(2)}]\sigma_\gamma^2P(im)\delta_{mn}\\
        \mathcal L^{(2)}_{ijmn}&\approx\frac{\mu_\gamma^2}{c^2N}f(im,jm)\delta_{mn}+\sigma_\beta^2\left(1+\tilde Q_i^{(1)}+\tilde Q_j^{(1)}+\sum_k(\tilde Q_{jk}^{(2)}+\tilde Q_{ik}^{(2)})\right)\delta_{mn}\\
        c\mathcal M^{(1)}_{ijmn}&\approx\mu_\gamma^2f(im,in)\delta_{ij}+[2\tilde R_i^{(1)}+2\tilde R_{ij}^{(2)}]\sigma_\gamma^2P(mi)\delta_{mn}\\
        \mathcal M^{(2)}_{ijmn}&\approx\frac{\mu_\gamma^2}{c^2N}f(mi,mj)\delta_{mn}+\sigma_\beta^2\left(1+\tilde R_i^{(1)}+\tilde R_j^{(1)}+\sum_k(\tilde R_{jk}^{(2)}+\tilde R_{ik}^{(2)})\right)\delta_{mn}
    \end{aligned}
\end{align}
Now consider the equilibrium equations additionally assuming $c_\gamma=0$
\begin{align}
    \begin{aligned}
        A_i^*&\approx\sum_j(1+Q_{ij}^{*(2)})(1+\beta^{*A}_j+\mu d_j^*+\epsilon_j^{*(1)}+\epsilon_j^{*(2)})\\
        P_i^*&\approx\sum_j(1+R_{ij}^{*(2)})(1+\beta^{*P}_j+\mu b_j^*+\eta_j^{*(1)}+\eta_j^{*(2)})\\
        \implies \frac{\delta A_i}{\delta 
        \epsilon_j}&=1+Q_{ij}^{*(2)}\\
        \frac{\delta P_i}{\delta \eta_j}&=1+R_{ij}^{*(2)}
    \end{aligned}
\end{align}
which by the definitions of the responses gives
\begin{align}
    \begin{aligned}
        \langle v\rangle Q_{ij}^{*(2)}&\approx\frac{\mu^2}{C}\Bigg\langle f(li,lj)(1+R_{ll}^{*(2)})\Bigg\rangle_l+\mu^2\Bigg\langle f(ki,li)(1+R_{lk}^{*(2)})\Bigg\rangle_{kl}\delta_{ij}\\
        \langle v\rangle R_{ij}^{*(2)}&\approx\frac{\mu^2}{C}\Bigg\langle f(il,jl)(1+Q_{ll}^{*(2)})\Bigg\rangle_l+\mu^2\Bigg\langle f(ik,il)(1+Q_{lk}^{*(2)})\Bigg\rangle_{kl}\delta_{ij}
    \end{aligned}
\end{align}
Again to first order this is simply the averages over the nestedness,
\begin{align}
    \begin{aligned}
        \langle v\rangle Q_{ij}^{*(2)}&\approx\frac{\mu^2}{C}\langle f(li,lj)\phi_l\rangle_l+\mu^2\langle f(ki,li)\phi_l\phi_k\rangle_{kl}\delta_{ij}\\
        \langle v\rangle R_{ij}^{*(2)}&\approx\frac{\mu^2}{C}\langle f(il,jl)\phi_l\rangle_l+\mu^2\langle f(ik,il)\phi_l\phi_k\rangle_{kl}\delta_{ij}
    \end{aligned}
\end{align}
A naïve approximation for the eigenvalues of the matrix can be given by just taking the averages of the blocks - we note that every term thus resolves into a number of summations over $f(ik,jk)$. Given that  $\sum_kf(ik,jk)=\Delta n_{ij}=\Delta f_{ij}{\rm min}(v_i,v_j)$ as defined in the previous section, the sum over all indices of these overlaps is again the object $\Delta X$, i.e. the difference between the value $X$ of the chosen network and one without any excess nestedness beyond what is obtained via structural constraints. This means that the variance of the terms of the matrix above are all proportional to $\Delta X$, which on setting $\sigma_\gamma=0$ gives
\begin{align}
    \begin{aligned}
        \sum_{ijmn} c\mathcal L^{(1)}_{ijmn}&\propto\mu_\gamma^2\sum_{imn}f(mi,ni)=\mu_\gamma^2\Delta {\rm X(plants)}\\
        \sum_{ijmn} c\mathcal M^{(1)}_{ijmn}&\propto\mu_\gamma^2\sum_{imn}f(im,in)=\mu_\gamma^2\Delta {\rm X(animals)}\\
         \sum_{ijmn} \mathcal L^{(2)}_{ijmn}&\propto\frac{\mu_\gamma^2}{c^2N}\Delta {\rm X(animals)}+\sigma_\beta^2(1+\frac{\mu^2}{c^2}\Delta {\rm X(animals/surviving \;plants)}+\frac{\mu^2}c\Delta {\rm X(surviving \;plants)})\\
         \sum_{ijmn} \mathcal M^{(2)}_{ijmn}&\propto\frac{\mu_\gamma^2}{c^2N}\Delta {\rm X(plants)}+\sigma_\beta^2(1+\frac{\mu^2}{c^2}\Delta {\rm X(plants/surviving \;animals)}+\frac{\mu^2}c\Delta {\rm X(surviving \;animals)})
    \end{aligned}
\end{align}
This means that the eigenvalues of that matrix move by values proportional to the eigenvalue of the 2x2 matrix composed of the values of the excess overlaps, with a certain spread related to the difference in the nestedness of the individual species. This leads us the same question as in the static case -- if the nestedness $\Delta F=\sum_{ij} f_{ij}-F_0=\sum _{ijk}f(ik,jk)/{\rm min}(v_i,v_j)$ is positive, what is the sign of $\Delta X$? Our conclusion thus remains the same: it might be possible to make a system more stable via nested interactions but this cannot be achieved by increasing overlaps blindly but rather careful strategising. On the other hand, increasing only one single overlap $f_{ij}$ is certain to reduce the stability of the system.

\end{document}